\documentclass[prc,aps,floatfix,twocolumn,groupedaddress,nofootinbib,showpacs,preprintnumbers,
amsmath,amssymb,amsfonts,widetable] {revtex4-1}
\usepackage{bm}
\usepackage{soul}
\usepackage{array}
\usepackage{lipsum}
\usepackage{relsize}
\usepackage{mathrsfs}
\usepackage{amssymb}
\usepackage{amsmath}
\usepackage{graphicx}
\usepackage[usenames]{color}
\usepackage{pslatex}
\usepackage{booktabs}
\usepackage{physics}
\usepackage{float}

\newcommand{\Lagrange}[1]{\mathscr{L}_{\raisebox{-0.5pt}{\scriptsize{#1}}}}
\newcommand{\RhoD}[2]{\rho_{\raisebox{-2pt}{\scriptsize{\!#1}\tiny{#2}}}}
\newcommand{\Fermi}[1]{{#1}_{\raisebox{-1pt}{\scriptsize{F}}}}
\newcommand{\Edens}{{\mathlarger{\varepsilon}}}

\newcommand{\rhozero}{\rho_{\raisebox{-1.0pt}{\tiny\!0}}}
\newcommand{\epszero}{\varepsilon_{\raisebox{-0.5pt}{\tiny 0}}}

\begin{document}
\title{Impact of tensor interactions and scalar mixing on covariant energy density functionals} 

\author{Marc Salinas and J. Piekarewicz}

\affiliation{Department of Physics, Florida State University, 
Tallahassee, FL 32306, USA}

\date{\today}

\begin{abstract}
The recent pioneering campaigns conducted by the Lead Radius Experiment (PREX) and the 
Calcium Radius Experiment (CREX) collaborations have uncovered major deficiencies in 
the theoretical description of some fundamental properties of atomic nuclei. Following a 
recent refinement to the isovector sector of covariant energy density functionals\,\cite{Reed:2023cap}, 
we present here additional improvements to the functional by including both tensor couplings 
and an isoscalar-isovector mixing term in the scalar sector. Motivated by the distinct surface 
properties of calcium and lead, we expect that the tensor terms that generate derivative 
couplings will help break the linear correlation between the neutron skin thickness of these 
two nuclei. Moreover, the addition of these new terms mitigates most of the problems identified 
in Ref.\cite{Reed:2023cap} in describing the properties of both finite nuclei and neutron stars. 
While significant progress has been made in reconciling the PREX-CREX results without 
compromising other observables, the final resolution awaits the completion of a proper 
calibration for this new class of functionals. We expect that powerful reduced basis methods 
used recently to create efficient emulators will be essential to accomplish this task.
\end{abstract}

\maketitle

\section{Introduction}
\label{Sec:Introduction}

More than three decades ago Donnelly, Dubach, and Sick suggested that measuring the parity violating 
asymmetry in elastic electron scattering may provide a model-independent determination of the neutron 
distribution of atomic nuclei\,\cite{Donnelly:1989qs}. Given that the nuclear charge distribution has been 
mapped with exquisite precision via parity conserving electron scattering, measuring the parity violating 
asymmetry offers the possibility for a model independent determination of the neutron skin thickness of 
a variety of nuclei. The neutron skin thickness is defined as the difference between the neutron and proton 
rms radii: $R_{\rm skin}\!=\!R_{n}\!-\!R_{p}$. Besides its intrinsic value as a fundamental nuclear structure 
observable, the neutron skin thickness of $^{208}$Pb has been shown to correlate strongly to the pressure 
of pure neutron matter in the vicinity of nuclear matter saturation density\,\cite{Brown:2000,Furnstahl:2001un}. 
This implies that an accurate measurement of the neutron skin thickness of $^{208}$Pb has also important 
implications on the structure of neutron stars\,\cite{Horowitz:2000xj,Horowitz:2001ya}. 

First proposed in 1999 and recently completed, the Lead Radius Experiment (PREX) at the Thomas Jefferson 
National Accelerator Facility (JLab) conducted two campaigns to determine the neutron skin thickness of 
$^{208}$Pb to be $R_{\rm skin}^{208} = 0.283 \pm 0.071\,\text{fm}$\,\cite{Abrahamyan:2012gp,Horowitz:2012tj,
Adhikari:2021phr}. By capitalizing on the strong correlation between $R_{\rm skin}^{208}$ and the slope of the 
symmetry energy $L$ at saturation density, a large value of $L\!=\!(106 \pm 37)\,\text{MeV}$ was 
reported\,\cite{Reed:2021nqk}. By itself, this result suggests that the symmetry energy is stiff, indicating a 
rapid increase in pressure with increasing density. Conceptually, the symmetry energy may be approximately
regarded as the energy required to convert symmetric nuclear matter into pure nuclear matter, although a more 
precise definition will be provided below. 

Shortly after the second PREX campaign ended, the Calcium Radius Experiment (CREX) at JLab was 
completed\,\cite{Adhikari:2022kgg}. Based on the strong correlation predicted in certain models 
between the neutron skin thickness of ${}^{48}$Ca and the corresponding neutron skin thickness of 
${}^{208}$Pb, it was anticipated that ${}^{48}$Ca will also display a relatively thick neutron skin. For
example, exploiting the strong correlation displayed by a certain class of covariant energy density
functionals (EDFs) between the neutron skin thickness of ${}^{48}$Ca and ${}^{208}$Pb, it was
predicted that $R_{\rm skin}^{\,48}\!=\!(0.229 \pm 0.035)\,{\rm fm}$\,\cite{Piekarewicz:2021jte}. Instead, 
the CREX collaboration reported the significantly smaller value of 
$R_{\rm skin}^{\,48}\!=\!(0.121\pm 0.035)$\,\cite{Adhikari:2022kgg}, where the error includes both 
experimental and theoretical uncertainties. Such unexpected result has caused a stir in the nuclear 
physics community. 

In an effort to find a resolution to the CREX-PREX ``dilemma'', several theoretical approaches have 
unsuccessfully attempted to reconcile these two measurements\,\cite{Hu:2021trw,Reinhard:2022inh,
Zhang:2022bni,Mondal:2022cva,Papakonstantinou:2022gkt,Yuksel:2022umn, Li:2022okx,Thakur:2022dxb,
Miyatsu:2023lki}. A common thread that has emerged from these studies suggests that accommodating 
the large value of $R_{\rm skin}^{208}$ within the constraints imposed by other nuclear observables---primarily 
the electric dipole polarizability and now also the neutron skin thickness of ${}^{48}$Ca---is enormously 
challenging. In fact, it has been suggested that due to the large error bars reported by the PREX collaboration, 
it is doubtful that the neutron skin thickness of ${}^{208}$Pb will provide any meaningful constraint on modern 
energy density functionals\,\cite{Reinhard:2022inh}. In this paper we explore the impact of expanding the
relatively simple isovector sector of an existing class of covariant EDFs.

Following the recent work of Reed and collaborators\,\cite{Reed:2023cap}, we include the isovector-scalar 
$\delta$-meson, as it was shown to have a significant impact on the neutron skin thickness of both $^{48}$Ca 
and $^{208}$Pb. However, due to the large isovector couplings adopted in such a work, the equation of state 
(EOS) of neutron rich matter became very stiff at the densities of relevance to the structure of neutron stars. 
This lead to predictions of large neutron star radii and tidal deformabilities, in disagreement with observations 
reported by both NICER\,\cite{Miller:2019cac,Riley:2019yda} and the LIGO-Virgo 
collaboration\,\cite{Abbott:PRL2017,Abbott:2018exr}. 
Besides its drawbacks at high densities, the strong isovector couplings of the models generate large density 
fluctuations in the nuclear interior that are not observed in the experimentally determined charge density. To 
mitigate some of these issues brought upon by the large couplings, we enlarge the isovector sector by 
including---in addition to the Yukawa coupling of the $\delta$-meson to the nucleon---a mixed isoscalar-isovector 
coupling in the scalar sector\,\cite{Li:2022okx,Thakur:2022dxb,Miyatsu:2023lki}, a quartic $\rho$-meson self-interaction 
term\,\cite{Mueller:1996pm}, as well as tensor couplings in the vector channel\,\cite{Rufa:1988zz,Typel:2020ozc}.

The main motivation for adding the tensor terms stems from the unique nucleon and meson spatial distributions 
in $^{48}$Ca and $^{208}$Pb. Given that both the proton and neutron distributions are smoother in lead than in 
calcium, this behavior is also reflected in the resulting meson fields. Consequently, one anticipates that the tensor 
interactions---which involve derivatives of the vector fields---will have a stronger effect in $^{48}$Ca than in $^{208}$Pb. 
Indeed, the hope is that such derivate (or ``gradient") terms will break the strong correlation between the neutron skins 
of calcium and lead. As we will show below, adding a tensor interaction in the $\omega$-meson channel indeed affects 
the neutron skin of calcium more than the one in lead. Unfortunately, improving the neutron skins comes at the expense of 
affecting other ground state properties, hence the need for an entirely new calibration of the model parameters. Although 
such a preliminary calibration will be attempted here, a full Bayesian analysis is deferred to a forthcoming work, where
we will bring to bear the full power of reduced basis methods\,\cite{Giuliani:2022yna}. In summary, the goal of this paper 
is threefold: (a) to showcase a new covariant EDF with a more sophisticated isovector sector, (b) to derive a new set of 
Kohn-Sham equations to re-calibrate the model parameters, and (c) to illustrate the positive impact that this new class of 
EDFs have in reproducing observables in both finite nuclei and neutron stars.

The paper has been organized as follows. In Sec.\ref{Sec:Formalism} we introduce the significantly augmented Lagrangian 
density and derive the associated mean field equations for both finite nuclei and infinite nuclear matter. We then proceed on
Sec.\ref{Sec:Results} to explore the individual effect of the various new terms and culminate by showing the overall impact
of the new Lagrangian density on the CREX-PREX observables as well as on the properties of neutron stars. We conclude
on Sec.\ref{Sec:Conclusions} with a summary of the important findings and prospects for a robust calibration of the model
parameters.
\vfill

\section{Formalism}
\label{Sec:Formalism}
\subsection{New Covariant Density Functional}

In the framework of covariant Density Functional theory (DFT), the underlying degrees of freedom are nucleons interacting via 
the exchange of mesons and the photon. In the particular version of covariant DFT used here\,\cite{Walecka:1974qa,Boguta:1977xi,
Serot:1984ey,Mueller:1996pm,Lalazissis:1996rd,Serot:1997xg,Horowitz:2000xj,Todd-Rutel:2005fa,Chen:2014sca,Chen:2014mza}, 
the effective Lagrangian density results from the sum of the three contributions: (a) a non-interacting term $\Lagrange{0}$, (b) a 
Yukawa term $\Lagrange{1}$ that couples the various mesons to the appropriate nucleon densities, and (c) a term $\Lagrange{2}$ 
containing both unmixed and mixed meson self-interactions. That is,
%%%%%%%%
\begin{equation}
 \Lagrange{} = \Lagrange{0} + \Lagrange{1} + \Lagrange{2}.
\end{equation}
%%%%%%%%

The free Lagrangian density includes the kinetic energy and mass term for the various constituents:
%%%%%%%%
\begin{widetext}
\begin{equation}
\begin{split}
 \Lagrange{0} &= \bar{\psi} \left( i \partial_\mu \gamma^{\,\mu} \!-\! M \right) \psi
                       + \left[\frac{1}{2} (\partial_\mu \phi)(\partial^\mu \phi) \!-\! \frac{1}{2} m_{\rm s}^2 \phi^2 \right] 
                       -\left[\frac{1}{4} V_{\mu \nu} V^{\mu \nu} \!-\! \frac{1}{2} m_{\rm v}^2 V_\mu V^\mu \right] \\
                       &+ \left[\frac{1}{2} (\partial_\mu \boldsymbol{\delta})( \partial^\mu \boldsymbol{\delta}) \!-\! 
                                 \frac{1}{2} m_\delta^2 (\boldsymbol{\delta} \cdot \boldsymbol{\delta})\right]
                       -\left[\frac{1}{4} {\bf b}_{\mu \nu} \cdot {\bf b}^{\mu \nu} \!-\! 
                               \frac{1}{2} m_\rho^2 ({\bf b}_{\mu} \cdot {\bf b}^{\mu} )\right]
                      -\frac{1}{4} F_{\mu\,\nu}F^{\mu\,\nu},        
\end{split}
\label{L0}
\end{equation}
\end{widetext}
%%%%%%%%
where $\gamma^{\;\mu}$ are the Dirac matrices and
%%%%%%%%
\begin{subequations}
\begin{align}
V_{\mu\,\nu} &= \partial_{\mu} V_{\nu} - \partial_{\nu} V_{\mu}, \\
{\bf b}_{\mu\,\nu} &= \partial_{\mu}{\bf b}_{\,\nu} - \partial_{\nu}{\bf b}_{\mu},  \\
F_{\!\mu\,\nu} &= \partial_{\mu} A_{\nu} - \partial_{\nu} A_{\mu}.  
\end{align}
\end{subequations}
%%%%%%%%
In the above expressions, $\psi$ is the isodoublet nucleon field, $A_{\mu}$ the photon field, and $\phi$, $V_{\mu}$, 
$\boldsymbol{\delta}$, and ${\bf b}_{\mu}$ represent the isoscalar-scalar $\sigma$-meson, the isoscalar-vector 
$\omega$-meson, the isovector-scalar $\delta$-meson, and the isovector-vector $\rho$-meson fields, respectively. 

In turn, the Yukawa component of the Lagrangian density is given by
%%%%%%%%
\begin{widetext}
\begin{eqnarray}
\Lagrange{1} =
\bar\psi \left[g_{\rm s}\phi   \!+\! g_{{}_{\delta}}{\boldsymbol{\delta}}\!\cdot\!\frac{{\boldsymbol{\tau}}}{2}
             \!-\! \left(g_{\rm v}\gamma^{\;\mu} +  f_{\rm v}\frac{\sigma^{\,\mu\nu}}{2M}\partial_{\nu}\right)\!V_{\mu}
             \!-\! \left(g_{\rho}\gamma^{\;\mu} +  f_{\rho}\frac{\sigma^{\,\mu\nu}}{2M}\partial_{\nu}\right)
                    {\bf b}_{\mu}\!\cdot\!\frac{\boldsymbol{\tau}}{2}
              \!-\! \frac{e}{2}\gamma^{\;\mu}A_{\mu}(1\!+\!\tau_{3})\right]\psi,
 \label{L1}
\end{eqnarray}
\end{widetext}
%%%%%%%%
where $\sigma^{\,\mu\nu}\!=\!i[\gamma^{\;\mu},\gamma^{\;\nu}]/2$, $\boldsymbol{\tau}$ is the vector containing the three 
Pauli matrices and $\tau_{3}$ is its $z$-component. Relative to the Lagrangian density introduced in Ref.\,\cite{Horowitz:2000xj}, 
the above Lagrangian density includes three additional Yukawa terms: one scalar $g_{{}_{\delta}}$ and two tensor couplings 
$f_{\rm v}$ and $f_{\rho}$. 

Lastly, we display the Lagrangian density containing both unmixed and mixed meson self interactions. That is, 
%%%%%%%%
\begin{widetext}
\begin{equation}
\Lagrange{2} =    - \frac{1}{3!} \kappa (g_{\rm s}\phi)^3 - \frac{1}{4!} \lambda (g_{\rm s} \phi)^4 
                            + \frac{1}{4!} \zeta (g_{\rm v}^2 V_\mu V^\mu)^2 
                            + \frac{1}{4!} \xi (g_\rho^2 \boldsymbol{b}_\mu \cdot \boldsymbol{b}^{\,\mu})^2  
                            - \Lambda_{\rm s} (g_\delta^2 \boldsymbol{\delta} \cdot \boldsymbol{\delta})(g_{\rm s}\phi)^2
                            + \Lambda_{\rm v} (g_\rho^2 \boldsymbol{b}_\mu \cdot \boldsymbol{b}^\mu)(g_{\rm v}^2 V_\mu V^\mu).
\end{equation}
\end{widetext}
%%%%%%%%
We observe that while minimal changes were made to the isoscalar sector\,\cite{Chen:2014sca}, the isovector sector was 
significantly enhanced. In addition to the ``standard" Yukawa coupling $g_{\rho}$ and the mixed term $\Lambda_{\rm v}$,
three additional parameters have been included: $g_{\delta}$, $\xi$, and $\Lambda_{\rm s}$.  However, it is important to note 
that $\Lagrange{2}$ incorporates only a subset of all possible meson self-interactions up to fourth order in the fields. Yet, the 
quartic self-interaction terms $\zeta$ and $\xi$ are crucial in softening the EOS at high densities. Indeed, it has been shown 
that it is possible to build different models that reproduce the same observed properties at normal nuclear densities, but which 
yield maximum neutron star masses that differ by more than one solar mass\,\cite{Mueller:1996pm}.

\subsection{Mean Field Approximation}
\label{Sec:MFA}

The field equations derived from the Lagrangian density given above may be solved exactly in the mean-field approximation. 
For static and spherically symmetric ground states, these equations have been known for several decades. However, when 
dealing with the extended Lagrangian density that includes both isovector-scalar and tensor terms, detailed derivations are 
scarce. To address this gap, we provide a comprehensive derivation for future reference, commencing with the meson field 
equations.

\subsubsection{Meson Field Equations}

In the mean-field approximation, the meson-field operators are replaced by their classical expectation values\,\cite{Serot:1997xg}, 
which for a static and spherically symmetric ground state imply the following simplifications:

%%%%%%%%%%%
\begin{subequations}
\begin{align}
\phi(x)  \rightarrow & \langle\phi(x)\rangle=\phi_{0}(r),\\
V^{\mu}(x)  \rightarrow & \langle V^{\mu}(x)\rangle=g^{\mu0}V_{0}(r),\\
\delta_{a}(x)  \rightarrow & \langle \delta_{a}(x)\rangle= \delta_{a3} \delta(r), \\
b_{a}^{\mu}(x)  \rightarrow & \langle b_{a}^{\mu}(x)\rangle=g^{\mu_{0}}\delta_{a3}b_{0}(r),\\
A^{\mu}(x)  \rightarrow & \langle A^{\mu}(x)\rangle=g^{\mu0}A_{0}(r).
\end{align}
\end{subequations}
%%%%%%%%%%

In turn, these classical meson fields satisfy Klein-Gordon equations containing both non-linear meson interactions and 
ground-state baryon densities as source terms. That is,
%%%%%%%%
\begin{widetext}
\begin{equation}
\begin{split}
    \left( m_{\rm s}^2 - \frac{\partial^2}{\partial r^2} - \frac{2}{r} \frac{\partial}{\partial r} \right) \Phi_0(r) &= g_{\rm s}^2 \left[\RhoD{s}{0}(r) -
    \frac{1}{2}\kappa\,\Phi_0^{2}(r) - \frac{1}{6} \lambda \Phi_{0}^{3}(r) - 2 \Lambda_{\rm s} \Delta_{0}^{2}(r) \Phi_0(r) \right],  \\
    \left( m_{\rm v}^2 - \frac{\partial^2}{\partial r^2} - \frac{2}{r} \frac{\partial}{\partial r} \right) W_0(r) &= g_{\rm v}^2 \left[\RhoD{v}{0}(r) -
    \frac{1}{2M} \frac{f_{\rm v}}{g_{\rm v}}\left( \frac{2}{r} + \frac{\partial}{\partial r} \right)\RhoD{t}{0}(r) -
    \frac{1}{6} \zeta W_{0}^{3}(r) - 2 \Lambda_{\rm v} B_{0}^{2}(r) W_0(r) \right], \\
    \left( m_\delta^2 - \frac{\partial^2}{\partial r^2} - \frac{2}{r} \frac{\partial}{\partial r} \right) \Delta_0(r) &= g_\delta^2 \left[ \frac{1}{2} \RhoD{s}{1}(r) - 
    2 \Lambda_{\rm s} \Phi_{0}^{2}(r) \Delta_0(r) \right],\\
    \left( m_{\rho}^2 - \frac{\partial^2}{\partial r^2} - \frac{2}{r} \frac{\partial}{\partial r} \right) B_0(r) &= g_{\rho}^2 \left[ \frac{1}{2} \RhoD{v}{1}(r) - 
    \frac{1}{4M} \frac{f_\rho}{g_\rho} \left( \frac{2}{r} + \frac{\partial}{\partial r} \right) \RhoD{t}{1}(r) -
    \frac{1}{6} \xi\,B_{0}^{3}(r) - 2 \Lambda_{\rm v} W_{0}^{2}(r) B_0(r) \right],
\end{split}
\label{KGEqs1}
\end{equation}
\end{widetext}
%%%%%%%%
where we have defined $\Phi_{0}\!=\!g_{\rm s}\phi_{0}$, $W_{0}\!=\!g_{\rm v}V_{0}$, $\Delta_{0}\!=\!g_{\delta}\delta_{0}$, and $B_{0}\!=\!g_{\rho}b_{0}$.
The subscripts $({\rm s,v,t})$ denote scalar, vector, and tensor Lorentz indices and ${\rm (0,1)}$ refer to isoscalar (proton-plus-neutron) and isovector 
(proton-minus-neutron) combinations, respectively. The Coulomb field $A_{0}$ satisfies the much simpler Poisson's equation with the proton vector density 
acting as the source term. That is, $\nabla^{2}A_{0}(r)\!=\!-e\rho_{p}(r)$.

\subsubsection{Dirac Equation}

Besides the various self-interacting terms appearing in Eq.(\ref{KGEqs1}), scalar, vector, and tensor densities also act as source terms for the meson fields.
These are given by
%%%%%%%%%%%
\begin{subequations}
\begin{align}
 \RhoD{s}{}(r) & = \sum_{n\kappa} \frac{2j+1}{4\pi\,r^{2}} \Big(g_{n\kappa}^{2}(r)-f_{n\kappa}^{2}(r)\Big), \\
 \RhoD{v}{}(r) & = \sum_{n\kappa} \frac{2j+1}{4\pi\,r^{2}} \Big(g_{n\kappa}^{2}(r)+f_{n\kappa}^{2}(r)\Big), \\
 \RhoD{t}{}(r) & = \sum_{n\kappa} \frac{2j+1}{4\pi\,r^{2}} \Big(2g_{n\kappa}(r)f_{n\kappa}(r)\Big),
\end{align}
\label{Rhos}
\end{subequations}
%%%%%%%%%%

where $g_{n\kappa}$ and $f_{n\kappa}$ are upper and lower components of the Dirac spinor. For spherically symmetric ground states,
these can be written in terms of spin-spherical harmonics as follows:
%%%%%%%%%
\begin{equation}
 {\cal U}_{\,n \kappa m}({\bf r}) = \frac {1}{r}
 \left( \begin{array}{c}
   \phantom{i}
   g_{n \kappa}(r) {\cal Y}_{+\kappa\,m}(\hat{\bf r})  \\
  if_{n \kappa}(r) {\cal Y}_{-\kappa\,m}(\hat{\bf r})
 \end{array} \right),
\label{Uspinor}
\end{equation}
%%
%%%%%%%%%
where $n$ is a radial quantum number and the spin-spherical harmonics ${\cal Y}_{\kappa\,m}(\hat{\bf r})$ are obtained by coupling the orbital 
angular momentum $l$ to the intrinsic nucleon spin to obtain a total angular momentum $j$ and magnetic quantum number $m$. As such, from
the generalized angular momentum quantum number $\kappa$, the total angular momentum $j$ and corresponding orbital angular momentum 
$l$ are obtained as follows:
%%%%%%%%%
\begin{equation}
 j = |\kappa| - \frac {1}{2} \;; \quad
 l = \begin{cases}
               \kappa\;,  & {\rm if} \; \kappa>0; \\
        -(1+\kappa)\;,  & {\rm if} \; \kappa<0, 
     \end{cases}
\end{equation}
%%%%%%%%%
Note that the orbital angular momentum of the upper and lower components differ by one unit, indicating that the orbital angular momentum is 
not a good quantum number---even for spherically symmetric ground states.

The Dirac Hamiltonian contains scalar, vector, and tensor terms of both isoscalar and isovector nature. That is,
%%%
\begin{equation} 
 \hat{H} = \bm{\alpha} \cdot \bm{p} + \beta\Big(M - S(r)\Big) + V(r) +i\,\bm{\gamma} \cdot\hat{\bm{r}}\,T(r),
\end{equation} 
%%%
where
%%%%%%%%%%%
\begin{subequations}
\begin{align}
S(r) & = \Phi_{0}(r) \pm \frac{1}{2} \Delta_{0}(r), \\
V(r) & = W_{0}(r) \pm \frac{1}{2} B_{0}(r) + eA_{0}(r)\frac{(1\!+\!\tau_{3})}{2},\\
T(r) & = \frac{1}{2M}\left[\frac{f_{\rm v}}{g_{\rm v}}\frac{dW_{0}(r)}{dr}\pm \frac{1}{2} \frac{f_{\rho}}{g_{\rho}}\frac{dB_{0}(r)}{dr}\right].
\end{align}
\end{subequations}
%%%%%%%%%%

Here the upper sign is for protons and the lower sign for neutrons. Now using the spherical symmetry of the problem, one derives a set of first order, 
coupled differential equations:
\begin{widetext}
\begin{subequations}
\begin{eqnarray}
  && \left(\frac{d}{dr}+\frac{\kappa}{r}+T(r)\right)g_{n\kappa}(r)-\Big(E+M-S(r)-V(r)\Big)f_{n\kappa}(r)=0,\\
  && \left(\frac{d}{dr}-\frac{\kappa}{r}-T(r)\right)f_{n\kappa}(r)+\Big(E-M+S(r)-V(r)\Big)g_{n\kappa}(r)=0.
\end{eqnarray}
\end{subequations}
\end{widetext}
%%%%%%%%
Finally, note that $g_{n\kappa}(r)$ and $f_{n\kappa}(r)$ satisfy the following normalization condition:
%%%%%%%%
\begin{equation}
    \int_{0}^{\infty} \Big(g_{n\kappa}^2(r) +f_{n\kappa}^2(r) \Big) dr=1.
 \label{norm}
\end{equation}
%%%%%%%%
From Eq.(\ref{Rhos}), this implies that both proton and neutron vector densities are conserved, namely, their corresponding integrals 
yield the number of protons $Z$ and the number of neutrons $N$, respectively. In contrast, neither the scalar nor the tensor densities 
are conserved.

\subsubsection{Ground State Properties}

Solving the field equations self-consistently results in single particle energies and associated Dirac orbitals from which ground state
densities and form factors may be extracted, so that they can then be compared against experimental data. The total energy of a 
system consisting of $Z$ protons and $N$ neutrons is obtained from the stress-energy-momentum tensor $T^{\alpha \beta}$ by 
integrating over all space. That is,
%%%%%%%%%
\begin{equation}
    E(Z,N) = \int T^{00}({\bf r}) d^{3}r \,.
\end{equation}
%%%%%%%%%

The total energy of the system may be expressed as a sum of its nucleonic $E_{\rm nuc}$ and mesonic contributions $E_{\rm mesons}$. 
The nucleonic contribution is simply obtained by adding the contribution from each single-particle orbital, namely,
%%%%%%%%%
\begin{equation}
    E_{\rm nuc} = \sum_{n \kappa}^{\rm occ} (2 j_{\kappa} + 1) E_{n \kappa}.
\end{equation}
%%%%%%%%%
For doubly-magic or semi-magic nuclei, all individual orbitals are completely filled. However, in cases in which the orbitals are partially filled,
a filling fraction must be specified. In turn, the contribution of the photon and the various meson fields to the total energy can be assessed 
in terms of their individual contributions. Following Ref.\,\cite{Giuliani:2022yna} we write 
%%%%%%%%%
\begin{equation}
    E_{\rm mesons} = 4\pi \int_0^\infty\!\Big( \Edens_{\sigma} \!+\! \Edens_\omega \!+\! \Edens_\delta \!+\! \Edens_\rho \!+\! 
                                                                   \Edens_\gamma \!+\! \Edens_{\sigma \delta} \!+\! \Edens_{\omega \rho} \Big) r^2 dr\,
\end{equation}
%%%%%%%%%
where the individual contributions are given by the following expressions:
%%%%%%%%%
\begin{widetext}
\begin{subequations}
  \begin{align}
        &\Edens_\sigma(r) = + \frac{1}{2}\Phi_0(r)\RhoD{s}{0}(r) - \frac{\kappa}{12}\Phi_0^3(r) - \frac{\lambda}{24}\Phi_0^4(r)\,,\\
        &\Edens_\omega(r)  = - \frac{1}{2}W_0(r)\RhoD{v}{0}(r) + \frac{1}{4M} \frac{f_{\rm v}}{g_{\rm v}} W_0(r) \!
                                          \left( \frac{2}{r} + \frac{\partial}{\partial r} \right)\!\RhoD{t}{0}(r) + \frac{\zeta}{24}W_0^4(r)\,,\\
        &\Edens_\delta(r)  = + \frac{1}{4} \Delta_0(r) \RhoD{s}{1}(r)\,,\\
        &\Edens_\rho(r)  = -\frac{1}{4} B_0(r)\RhoD{v}{1}(r) + \frac{1}{8M} \frac{f_\rho}{g_\rho} B_0(r)\! 
                                    \left( \frac{2}{r} + \frac{\partial}{\partial r} \right)\!\RhoD{t}{1}(r) + \frac{\xi}{24} B_0^4(r)\,,\\
        &\Edens_\gamma(r)  = -\frac{1}{2} eA_0(r) \rho_{\rm vp}(r) \,,\\
        &\Edens_{\sigma \delta}(r)  = - \Lambda_{\rm s} \Phi_0^2(r) \Delta_0^2(r) \,,\\
        &\Edens_{\omega \rho}(r)  =  + \Lambda_{\rm v} W_0^2(r) B_0^2(r) \,.
    \end{align}
\end{subequations}
\end{widetext}
%%%%%%%%%
Relative to the expressions given in Ref.\,\cite{Giuliani:2022yna}, we find several additional contributions due to the inclusion of:
(a) the isovector $\delta$ meson, with both a Yukawa coupling to the nucleon and a coupling to the isoscalar $\sigma$-meson, 
(b) a quartic self-interaction term for the $\rho$-meson, and (c) both isoscalar and isovector tensor terms. Finally, the total 
binding energy of the system is obtained by removing the nucleon rest mass and adding a center-of-mass correction.
That is,
%%%%%%%%%
\begin{equation}
    B(Z,N) = E(Z,N) + E_{\rm cm} - (Z+N)\,M,
\end{equation}
%%%%%%%%%
where we have adopted a center-of-mass correction obtained from a harmonic oscillator approximation\,\cite{Bender:2003jk}:
%%%%%%%%%
\begin{equation}
   E_{\rm cm} = -\frac{3}{4}\,41\,A^{-1/3}\,{\rm MeV}.
\end{equation}
%%%%%%%%%

Besides the single particle Dirac orbitals, one can extract ground state scalar, vector, and tensor densities, as defined in Eqs.(\ref{Rhos}). 
In particular, charge and weak-charge densities may be compared directly to experiment by folding the self-consistently obtained proton 
and neutron vector densities with suitable nucleon form factors that account for the finite nucleon size. Particularly useful to compare 
against experiment are a few of the low moments of the spatial distribution which, using the charge distribution as an example, are given 
by
%%%%%%%%%%
\begin{subequations}
\begin{align}
 R_{\rm ch}^{\,2} & = \frac{1}{Z} \int r^2 \RhoD{ch}{}({\bf r}) d^3{r} \rightarrow \frac{3}{5}c^{2} + \frac{7}{5}(\pi a)^{2} \,,\\
 R_{\rm ch}^{\,4} & = \frac{1}{Z} \int r^4 \RhoD{ch}{}({\bf r}) d^3{r} \rightarrow \frac{3}{7}c^{4} + \frac{18}{7}(\pi a)^{2}c^{2} + \frac{31}{7}(\pi a)^{4} \,.
\end{align}
\label{SFMoments}
\end{subequations}
%%%%%%%%%%
The arrows in the above expressions indicate exact expressions in the case that the density may be represented as a two-parameter 
symmetrized Fermi distribution\,\cite{Sprung:1997,Piekarewicz:2016vbn}, with $c$ defined as  the half-density radius and $a$ as the surface 
diffuseness. Thus, for a heavy nucleus such as  ${}^{208}$Pb, two measurements of the form factor may be sufficient to accurately describe 
its spatial distribution of charge and weak-charge\,\cite{Piekarewicz:2016vbn}.

%%%%%%%%%%%%%%%%%%%%%%%%%%%%%%%%%%%%%%%%%%%%%%%%%%%%%%%
%%%%%%%%%%%%%%%%%%%%%%%%%%%%%%%%%%%%%%%%%%%%%%%%%%%%%%%
\bigskip
\subsection{Infinite Nuclear Matter}

The aim of a robust energy density functional is to describe nuclear observables ranging from the properties of finite nuclei to the structure of
neutrons stars. To do so, one must supplement the framework provided in the previous section with one that computes the equation of state
of infinite nuclear matter as a function of the total baryon density of the system $\rho\!=\!\rho_{\rm p}\!+\!\rho_{\rm n}$ and the neutron proton 
asymmetry $\alpha\!=\!(\rho_{\rm p}\!-\!\rho_{\rm n})/\rho$. For isolated neutron stars, the characteristic Fermi temperature is much larger than 
the actual temperature of the star, so it is appropriate to construct a zero-temperature equation of state. In this way, all intensive quantities in
the system, such as the energy density and the pressure, depend only on $\rho$ and $\alpha$. Hence, by varying the neutron proton asymmetry, 
one can account for the equation of state of symmetric nuclear matter ($\alpha\!=\!0$) and pure neutron matter ($\alpha\!=\!1$). In turn, the EOS 
of neutron star matter is obtained by demanding chemical (or ``beta") equilibrium that consists on finding the absolute ground state of the system
at a given density. That is, the EOS in chemical equilibrium is obtained by computing the optimal value of $\alpha$ that minimizes the total energy 
of the system at a given density. 

In the stellar core where the system is spatially uniform, the meson fields become independent of all spatial coordinates so all derivatives
vanish in this limit. Relative to Eqs.(\ref{KGEqs1}), the meson field equations simplify considerably; one obtains:
%%%%%%%%
\begin{widetext}
\begin{equation}
\begin{split}
    m_{\rm s}^2 \Phi_0 &= g_{\rm s}^2 \left[\RhoD{s}{0} - \frac{1}{2}\kappa\,\Phi_0^{2} - \frac{1}{6} \lambda \Phi_{0}^{3} 
                                      - 2 \Lambda_{\rm s} \Delta_{0}^{2} \Phi_0 \right],  \\
    m_{\rm v}^2 W_0 &= g_{\rm v}^2 \left[\RhoD{v}{0} - \frac{1}{6} \zeta W_{0}^{3} - 2 \Lambda_{\rm v} B_{0}^{2}W_0 \right], \\
    m_\delta^2\Delta_0 &= g_\delta^2 \left[ \frac{1}{2} \RhoD{s}{1}- 2 \Lambda_{\rm s} \Phi_{0}^{2} \Delta_0 \right],\\
    m_{\rho}^2 B_0 &= g_{\rho}^2 \left[ \frac{1}{2} \RhoD{v}{1} - \frac{1}{6} \xi\,B_{0}^{3} - 2 \Lambda_{\rm v} W_{0}^{2}B_0\right].
\end{split}
\label{KGEqs2}
\end{equation}
\end{widetext}
%%%%%%%%

Given that the tensor terms appearing in Eq.(\ref{L1}) involve derivatives of the vector fields, these terms vanish in infinite nuclear matter, so their 
impact on the EOS is indirect, namely, only through the calibration of the entire set of parameters to the properties of finite nuclei. 

We now proceed to compute the EOS of infinite nuclear matter---the relation between the energy density and the pressure---in terms of the 
stress-energy-momentum tensor $T^{\alpha \beta}$. That is,
%%%%%%%%
\begin{widetext}
\begin{equation}
\begin{split}
\Edens & = \Edens(\RhoD{p}{}) + \Edens(\RhoD{n}{}) + \frac{1}{2} \frac{m_{\rm s}^2}{g_{\rm s}^2} \Phi_0^2 + \frac{1}{2} \frac{m_{\rm v}^2}{g_{\rm v}^2} W_0^2 + \frac{1}{2} \frac{m_\delta^2}{g_\delta^2} \Delta_{0}^2  + \frac{1}{2} \frac{m_\rho^2}{g_\rho^2} B_0^2 
               + \frac{\kappa}{6}\Phi_{0}^3 + \frac{\lambda}{24}\Phi_{0}^4 + \frac{\zeta}{8} W_0^4 + \frac{\xi}{8} B_0^4 + \Lambda_{\rm s} \Delta_0^2 \Phi_{0}^2 + 3 \Lambda_{\rm v} B_0^2 W_0^2\,,  \\
P & = P(\RhoD{p}{}) + P(\RhoD{n}{}) - \frac{1}{2} \frac{m_{\rm s}^2}{g_{\rm s}^2} \Phi_0^2 + \frac{1}{2} \frac{m_{\rm v}^2}{g_{\rm v}^2} W_0^2 - \frac{1}{2} \frac{m_\delta^2}{g_\delta^2} \Delta_{0}^2 + \frac{1}{2} \frac{m_\rho^2}{g_\rho^2} B_0^2 
       - \frac{\kappa}{6}\Phi_{0}^3 - \frac{\lambda}{24}\Phi_{0}^4 + \frac{\zeta}{24} W_0^4 + \frac{\xi}{24} B_0^4 - \Lambda_{\rm s} \Delta_0^2 \Phi_{0}^2+ \Lambda_{\rm v}B_0^2 W_0^2\,.
\end{split}
\label{EvsP1}
\end{equation}
\end{widetext}
%%%%%%%%
The nucleonic contributions to the energy density may be computed from the standard expression for a free Fermi gas:
%%%%%%%%
\begin{widetext}
\begin{equation}
\begin{split}
\Edens(\rho)  & = \frac{1}{\pi^2} \int_0^{k_{\rm F}} k^2 \sqrt{k^2 + M^{\,2}} dk 
                         = \frac{M^{4}}{8\pi^{2}}\Big[\Fermi{x}\Fermi{y}(\Fermi{x}^{2}+\Fermi{y}^{2})-\ln(\Fermi{x}+\Fermi{y})\Big], \\
P(\rho)  & = \frac{1}{3\pi^2} \int_0^{k_{\rm F}} \frac{k^4}{\sqrt{k^2 + M^{\,2}}} dk    
                =  \frac{M^{4}}{8\pi^{2}}\Big[\frac{2}{3}\Fermi{x}^{3}\Fermi{y}-\Fermi{x}\Fermi{y}+\ln(\Fermi{x}+\Fermi{y})\Big],                 
\end{split}
\label{EvsP2}
\end{equation}
\end{widetext}
%%%%%%%%
where the scaled Fermi momentum, Fermi energy, and number density are given by
%%%%%%%%
\begin{equation}
 \Fermi{x}\!\equiv\!\frac{\Fermi{k}}{M}, \hspace{5pt}\Fermi{y}\!\equiv\!\sqrt{1+\Fermi{x}^{2}}, 
                \hspace{5pt}{\rm and}\hspace{3pt}\rho=\frac{\Fermi{k}^{3}}{3\pi^{2}}.
\end{equation}
%%%%%%%%
It is important to note that the formalism implemented here is such that both Eq.(\ref{EvsP1}) and Eq.(\ref{EvsP2}) are fully 
consistent with the Hugenholtz-Van Hove theorem\,\cite{Hugenholtz:1958zz,Satpathy:1983}, which demands that the pressure,
energy density, baryon density, and Fermi energy satisfy the relation $P+\Edens=\rho\Fermi{E}$.

We conclude this section by identifying some bulk properties of infinite nuclear matter that offer valuable insights into the
underlying dynamics. To start, we expand the energy per nucleon in even powers of the neutron-proton asymmetry $\alpha$.
That is,
%%%%%%%%%
\begin{equation}
  \frac{E}{A}(\rho,\alpha) -\!M \equiv E(\rho,\alpha) = E_{\rm SNM}(\rho) + \alpha^{2} S(\rho) + {\cal O}(\alpha^{4}) \,,
 \label{EOS1}
\end {equation}
%%%%%%%%%
where $E_{\rm SNM}(\rho)\!=\!E(\rho,\alpha\!=\!0)$ is the energy per nucleon of symmetric nuclear matter (SNM) and 
the symmetry energy $S(\rho)$ represents the first-order correction to the symmetric limit. Note that no odd powers of 
$\alpha$ appear in the expansion since in the absence of electroweak interactions the nuclear force is assumed to be 
isospin symmetric. By the same token, the energy of pure neutron matter may be written as
%%%%%%%%%
\begin{equation}
  E_{\rm PNM}(\rho) = E(\rho,\alpha\!=\!1) = E_{\rm SNM}(\rho) + S(\rho)  + \ldots
 \label{EOS2}
\end {equation}
%%%%%%%%%
Although there is \emph{a priori} no reason to neglect the higher order terms in Eq.(\ref{EOS2}), for certain classes of 
density functionals the symmetry energy seems to be a good approximation to the energy cost required to convert 
symmetric nuclear matter into pure neutron matter, namely, 
%%%
\begin{equation}
 S(\rho)\!\approx\! \Big(E_{\rm PNM}(\rho)-E_{\rm SNM}(\rho)\Big)\,.
 \label{SymmE}
\end {equation}
%%%
However, while the validity of this relation is easily verified when both protons and neutrons behave as non-interacting 
Fermi gases\,\cite{Piekarewicz2016}, there is no guarantee that the approximation will remain valid in the presence of 
interactions. 

The separation of the energy per nucleon as in Eq.(\ref{EOS1}) is useful because symmetric nuclear matter is 
sensitive to the isoscalar sector of the density functional which is well constrained by the properties of stable 
nuclei. In contrast, the nuclear symmetry energy quantifies the cost of having excess neutrons (or protons) and
it is therefore highly sensitive to the less constrained isovector sector. Further, it is useful to characterize the EOS 
in terms of a few bulk parameters defined at saturation density. Although not directly measurable, these bulk
parameters are often strongly correlated to some fundamental experimental observables. By performing a Taylor 
series expansion around nuclear matter saturation density $\rhozero$, one obtains\,\cite{Piekarewicz:2008nh}:
%%%
\begin{subequations}
\begin{align}
 & E_{\rm SNM}(\rho) = \epszero + \frac{1}{2}K_{0}\,x^{2}+\ldots ,\label{EandSa}\\
 & S(\rho) = J + L\,x + \frac{1}{2}K_{\rm sym}x^{2}+\ldots ,\label{EandSb}
\end{align} 
\label{EandS}
\end{subequations}
%%%
where $x\!=\!(\rho-\rhozero)\!/3\rhozero$ is a dimensionless parameter that quantifies the deviations of the 
density from its value at saturation. Here $\epszero$ and $K_{0}$ represent the energy per nucleon and the 
incompressibility coefficient of SNM. Notably, the linear term in this expression is absent because the pressure 
of symmetric nuclear matter vanishes at saturation. When examining the corresponding values for the symmetry 
energy, it is conventional to denote these quantities by $J$ and $K_{\text{sym}}$. However, unlike the case of 
symmetric nuclear matter, the slope of the symmetry energy $L$ does not vanish. In fact, assuming the validity 
of the so-called parabolic approximation outlined in Eq.(\ref{SymmE}), the slope of the symmetry energy is 
directly proportional to the pressure of pure neutron matter at saturation density, namely, $P_{0} \approx \rhozero L/3$.

%%%%%%%%%%%%%%%%%%%%%%%%%%%%%%%%%%%%%%%%%%%%%%%%%%%%%
%%%%%%%%%%%%%%%%%%%%%%%%%%%%%%%%%%%%%%%%%%%%%%%%%%%%%
\section{Results}
\label{Sec:Results}

Having expressed various physical observables in terms of the extended set of model parameters, we now proceed to calibrate 
the new functionals. We perform a modest fit by optimizing a likelihood function written in terms of a traditional chi-square 
objective function $\mathcal{L}\!\propto\!\exp (-\chi^2/2)$. Experimental ground-state energies and charge radii for $^{40}$Ca, 
$^{48}$Ca, and $^{208}$Pb are used in the calibration procedure---as well as the new information on the weak-skin form factor
extracted from the PREX\,\cite{Adhikari:2021phr} and CREX\,\cite{Adhikari:2022kgg} campaigns. The weak-skin form 
factor is defined as follows\,\cite{Thiel:2019tkm}:
%%%%%%%%
\begin{equation}
 F_{\rm Wskin}(q)\!\equiv\!F_{\text{ch}}(q)\!-\!F_{\text{wk}} (q) \xrightarrow[q \rightarrow 0]{} 
%      \frac{q^{2}}{6}\left(R_{\rm wk}^{2}-R_{\rm ch}^{2}\right) = 
      \frac{q^{2}}{6}\left(R_{\rm wk}\!+\!R_{\rm ch}\right)R_{\rm Wskin},
 \label{Fwskin}
\end{equation}
%%%%%%%%
where $F_{\text{ch}}$ and $F_{\text{wk}}$ are the charge and weak form factors measured at the momentum transfer of the
experiment, namely, $q\!=\!0.398\,{\rm fm}^{-1}$ for PREX and $q\!=\!0.873\,{\rm fm}^{-1}$ for CREX. In turn, the weak skin is
defined in terms of the corresponding experimental radii as $R_{\rm Wskin}\!\equiv\!R_{\rm wk}\!-\!R_{\rm ch}$. Relative to the 
better known neutron skin thickness $R_{\rm skin}\!=\!R_{\rm n}\!-\!R_{\rm p}$, the weak skin differs from it by incorporating 
corrections coming from the finite nucleon size\,\cite{Horowitz:2012we}. 

Due to the high dimensionality of the parameter space and the computational demands imposed by self-consistent, mean-field 
calculations, we take a slightly unorthodox approach to the calibration procedure. We start by defining the seven model 
parameters that enter into the likelihood function as follows: ${\bf p}\!=\!\{m_{\text{s}}, \epszero, \rhozero, M^*, K_0, f_{\rm v}, f_\rho\}$.
This set of parameters is varied within a pre-defined range while leaving $m_{\text{v}}, m_\rho, m_\delta, \tilde{J}, L, K_{\rm sym}, 
\zeta, \xi, \Lambda_s$ fixed; here $\tilde{J}$ is the value of the symmetry energy at a density  
$\tilde{\rho}\!=\!2\rhozero/3\!\approx\!0.1\,{\rm fm}^{-1}$. It is important to note that there is a one-to-one correspondence between 
four of the bulk parameters characterizing symmetric nuclear matter, namely, $\epszero, \rhozero, M^*, K_0$ and the four isoscalar 
model parameters $g_{\rm s}^{2}/m_{\rm s}^{2}$, $g_{\rm v}^{2}/m_{\rm v}^{2}$, $\kappa$, and $\lambda$\,\cite{Glendenning:2000}. 
Beyond the isoscalar sector, it was shown that the two isovector parameters $g_{\rho}^{2}/m_{\rho}^{2}$ and $\Lambda_{\rm v}$ may 
also be determined from knowledge of two quantities of central importance to the symmetry energy, namely, $J$ (or $\tilde{J}$) and 
$L$\,\cite{Chen:2014sca}. These relations have now been extended here by using $J$, $L$, and $K_{\rm sym}$ to also determine 
$g_{\delta}^{2}/m_{\delta}^{2}$. The connection between bulk properties and model parameters has proved to be extremely useful, 
both in developing an intuitive picture of the parameter space as well as on improving the convergence of the self-consistent 
calculations.

Once the choice of parameters has been made, each relevant observable ${\cal O}({\bf p})$ is sampled at three different points 
for each of the seven parameters in ${\bf p}$, thereby creating independent parabolic approximation to ${\cal O}({\bf p})$ along 
each of the seven directions. That is, given a current location in parameter space ${\bf p}_{{}_{0}}$, the parabolic approximation
to ${\cal O}({\bf p})$ is given by
%%%%%%%%
\begin{equation}
 {\cal O}({\bf p})= {\cal O}({\bf p}_{{}_{0}}) + \sum_{n=1}^{7}\Big[a_{n}+b_{n}(p_{n}-c_{n})^{2}\Big],
\end{equation}
%%%%%%%%
where $a_{n}$, $b_{n}$, and $c_{n}$ are obtained from sampling the observable at three different points in every direction.
The parabolic approximation yields a gaussian likelihood function that can be optimized using algorithm packages available in 
Mathematica, such as Random Search, Nelder Mead, and/or Simulated Annealing. The parameter set ${\bf p}$ that optimizes the 
likelihood function is then re-sampled and the procedure is repeated until it converges to the desired accuracy.

The main disadvantages of using the method just described is the lack of quantification of errors and correlations, as well as 
having to limit the model space by fixing certain parameters so the dimensionality of the space becomes manageable. Monte Carlo 
sampling methods, such as was done in our recent publication\,\cite{Salinas:2023nci}, might be more suitable in the future with the 
emergence of Reduced Basis Methods (RBM) techniques\,\cite{Bonilla:2022rph,Melendez:2022kid,Giuliani:2022yna} that will be 
explored in a future work.

\begin{center}
\begin{figure}[h]
\centering
\includegraphics[width=0.48\textwidth]{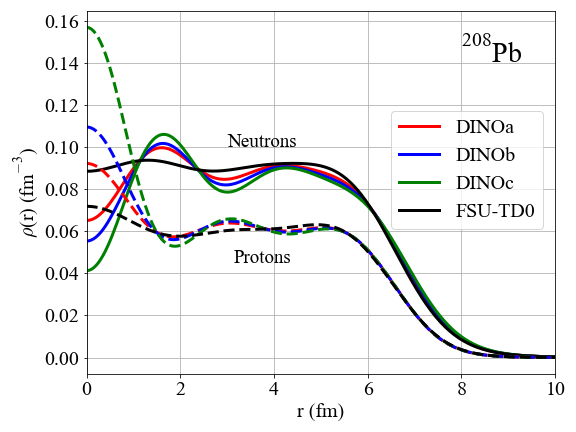}
\caption{Point proton (dashed lines) and neutron (solid lines) densities of $^{208}$Pb as predicted by the three DINO models\,\cite{Reed:2023cap}, alongside predictions from the FSU-TD0 model introduced in this work (see Table\,\ref{Table4}). The FSU-TD0 model eliminates the large oscillations in the nuclear interior by trading the $\delta$-meson term
in the DINO models in favor of a $\rho$-meson tensor interaction.} 
\label{Fig0}
\end{figure}
\end{center}

To set the stage and motivate the need for a new set of covariant EDFs, we display in Fig.\ref{Fig0} point proton and neutron densities for $^{208}$Pb as predicted by the recently calibrated DINO models\,\cite{Reed:2023cap}. Also shown are predictions from one of the newly extended models introduced in this work (FSU-TD0) and discussed in great detail in the following sections. Although the addition of a third isovector parameter ($g_{\delta}$) enabled the DINO models to reproduce the CREX-PREX results at the 67\% confidence level, this success came with a very heavy price, namely, generating large unphysical oscillations in the nuclear interior. As we show in the following sections and clearly illustrated in the figure, the inclusion of tensor terms in the calibration of the FSU-TD functionals becomes instrumental in removing these large oscillations.  

%%%%%%%%
%%%%%%%%

In the following sections we will start by examining---first without re-tuning the model parameters---the individual impact of some of 
the new terms added to the Lagrangian density. We will then proceed to re-tune the parameters using ground-state properties of
$^{40}$Ca, $^{48}$Ca, and $^{208}$Pb, and will end the section by showcasing some of the results obtained with the newly 
calibrated models.

%%%%%%%%%%%%%%%%%%%%%%%%%%%%%%%%%%%%%%
%%%%%%%%%%%%%%%%%%%%%%%%%%%%%%%%%%%%%%
\subsection{Tensor coupled $\rho$-meson}

\vspace{-12pt}
\label{Sec:rhotensor}
To our knowledge, incorporating tensor interactions to the relativistic Lagrangian was first reported in a work by Rufa et al.\cite{Rufa:1988zz};
for a more recent reference see Ref.\,\cite{Typel:2020ozc} and references contained therein. 
The Lagrangian density introduced in that work included Yukawa couplings to the $\sigma$, $\omega$, $\rho$, and photon fields, cubic and 
quartic scalar self-interactions, and tensor coupling terms involving the two vector mesons. A particularly important result drawn from such
study concluded that the $\omega$-tensor interaction increases the effective nucleon mass while maintaining a robust spin-orbit splitting. In 
contrast, the authors found that the $\rho$-tensor interaction offers very little to justify its inclusion. In this work we demonstrate its significant 
impact on the charge density as well as on the weak skin. However, we note that it is difficult to cleanly isolate the impact of each individual term 
on a specific observable because of the strong correlation among model parameters.

\begin{center}
\begin{figure}[H]
\centering
\includegraphics[width=0.48\textwidth]{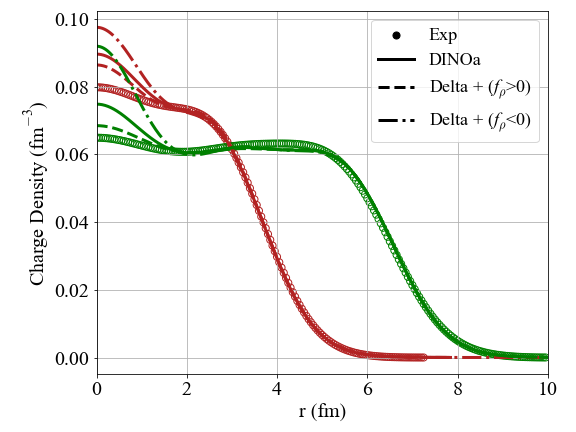}
\caption{Charge density of $^{48}$Ca (red) and $^{208}$Pb (green) as predicted by the DINOa model\,\cite{Reed:2023cap} with and without
the inclusion of a tensor-coupled $\rho$-meson. For $f_\rho\!=\!-4$ (dot-dashed line), the density fluctuations in the nuclear interior increase. In contrast, a positive $f_\rho\!=\!10$ (dashed line) mitigates the charge density fluctuations in the interior, thereby improving the 
agreement with experiment (circled points)\,\cite{DeJager:1987qc}. Predictions from the original DINOa model are depicted with a solid line.}
\label{Fig1}
\end{figure}
\end{center}

We start by considering one of the ``DINO" models introduced in Ref.\cite{Reed:2023cap}. By adding the isovector $\delta$-meson, the DINO 
models provide an extension of the FSU-like models\,\cite{Todd-Rutel:2005fa} in the hope of mitigating the PREX-CREX 
discrepancy\,\cite{Adhikari:2022kgg}. Given that the curvature of the symmetry energy $K_{\rm sym}$ was believed to hold the key to elucidate 
the source of the discrepancy, a third isovector parameter ($g_{\delta}^{2}$) was added in order to tune $J$, $L$, and $K_{\rm sym}$. 
In Fig.\ref{Fig1} we examine the predictions of the DINOa model\,\cite{Reed:2023cap} on the charge density of $^{48}$Ca and $^{208}$Pb with 
and without the influence of the $\rho$-tensor coupling $f_{\rho}$. We note that for negative $f_\rho$, the already large density fluctuations in the 
nuclear interior predicted by the DINO model are further exacerbated. Instead, a positive $f_\rho$ reduces the density fluctuations at the center of 
the nucleus bringing the predictions into closer agreement with experiment. We also display in Table\,\ref{Table1} the impact of $f_\rho$ on some
ground-state observables of relevance to PREX and CREX---without any re-tuning of the model parameters. As alluded in Ref.\,\cite{Rufa:1988zz},
the impact of the $\rho$-tensor interaction on the properties of $^{208}$Pb is indeed modest. With or without the addition of $f_\rho$, the DINOa
predictions for the neutron skin thickness remains within 1.5 standard deviations from the PREX central value. This, however, is not the case 
for $^{48}$Ca where the changes in the neutron skin thickness are significant. Let us restate that no effort has been made so far in re-calibrating
the model parameters. Suffices to say that the inclusion of the $\rho$-tensor interaction---a derivative coupling that enhances surface effects---is
worth incorporating into these new covariant EDFs.

%%%%%%%%
%%%%%%%%

%%%%%%%%  Table I  %%%%%%%
\begin{table}[h]
\begin{center}
\begin{tabular}{|l||c|c|c|c|c|}
 \hline\rule{0pt}{2.25ex}   
 \!\!Model (${}^{208}$Pb)& $B/A (\rm MeV)$ & $R_{\rm ch} (\rm fm)$ & $F_{\rm ch}\!-\!F_{\rm wk}$ & $R_{n}\!-\!R_{p} (\rm fm)$ \\
 \hline
 \hline\rule{0pt}{2.25ex} 
 \!\!DINOa       & 7.864 &  5.508 & 0.0264 & 0.1753 \\ 
     DINOa+($f_\rho\!>\!0$)  & 7.852  & 5.510  & 0.0268 & 0.1794  \\
     DINOa+($f_\rho\!<\!0$)  & 7.870  & 5.504  & 0.0263 & 0.1742  \\ 
 \hline\rule{0pt}{2.25ex} 
     Experiment   &  7.867 &    5.501(1) &  0.041(13)& 0.283(71) \\
 \hline                                                                                                 
 \hline\rule{0pt}{2.25ex}   
\!\!Model (${}^{48}$Ca)& $B/A (\rm MeV)$ & $R_{\rm ch} (\rm fm)$ & $F_{\rm ch}\!-\!F_{\rm wk}$ & $R_{n}\!-\!R_{p} (\rm fm)$ \\
 \hline
 \hline\rule{0pt}{2.25ex} 
 \!\!DINOa       &  8.674 &  3.468 & 0.0334 & 0.1002 \\ 
     DINOa+($f_\rho\!>\!0$)  & 8.638  & 3.463  & 0.0378 & 0.1277 \\
     DINOa+($f_\rho\!<\!0$)  & 8.695  & 3.463  & 0.0311 & 0.0871  \\ 
 \hline\rule{0pt}{2.25ex} 
     Experiment     & 8.667 & 3.477(2) & 0.0277(55) & 0.121(35) \\
 \hline                                                                                                 
\end{tabular}
\caption{Ground-state observables for  $^{208}$Pb and $^{48}$Ca as predicted by the original DINOa model\,\cite{Reed:2023cap}  
              and the DINOa model supplemented by a tensor-coupled $\rho$-meson, with $f_\rho\!=\!10$ and $f_\rho\!=\!-4$. 
              No attempt has been made at this point to re-calibrate the model parameters.}
\label{Table1}
\end{center}
\end{table}
%%%%%%%%%%%%%%%%%%%%
%For the new class of functionals we will calibrate, we are clearly left with two choices when trying to resolve the charge density fluctuations that result from the inclusion of the $\delta$-meson. We can set $f_\rho>0$, which will guarantee a reduction in the large charge densities we see at the center of the nucleus. We can then try and counteract the negative impact on the weak skins through the modification of other parameters. The other option, which seems to be more favorable in the minimization procedure, is to calibrate these new models using a negative $f_\rho$ coupling to improve the weak skins while simultaneously reducing the large isovector couplings. This should be able to smoothen out the charge density fluctuations, while still reproducing CREX/PREX results to the same extent as the new DINO models. After the calibration procedure, the hope is that the adjustment of other couplings would restore the binding energies and charge radii to their experimental values, while maintaining the effect on the skins.
%%%%%%%%%%%%%%%%%%%%%%%%%%%%%%%%%%%%%%%%%%%%%%%%%%%%%%%%%%%%%%%%%%%
%%%%%%%%%%%%%%%%%%%%%%%%%%%%%%%%%%%%%%%%%%%%%%%%%%%%%%%%%%%%%%%%%%%
% FIXED ABOVE

\subsection{Tensor coupled $\omega$-meson}
\label{Sec:omega tensor}

The main motivation for adding an isoscalar tensor term stems from the difference in the spatial dependence of the $\omega$-meson field for 
$^{48}$Ca and $^{208}$Pb. Given the derivative nature of the tensor coupling, one expects a more significant effect for $^{48}$Ca where the
spatial fluctuations are larger than for $^{208}$Pb where the $\omega$-meson field is mostly uniform throughout the nuclear interior. When adding the
$\omega$-tensor coupling $f_{\rm v}$, we witness the anticipated mass dependence, particularly on observables sensitive to the nuclear surface. 
As in Table\,\ref{Table1}, we display in Table\,\ref{Table2} the quantitative impact of $f_{\rm v}$ on a few selected ground-state observables.

%%%%%%%%  Table II  %%%%%%%
\begin{table}[H]
\begin{center}
\begin{tabular}{|l||c|c|c|c|c|}
 \hline\rule{0pt}{2.25ex}   
 \!\!Model (${}^{208}$Pb)& $B/A (\rm MeV)$ & $R_{\rm ch} (\rm fm)$ & $F_{\rm ch}\!-\!F_{\rm wk}$ & $R_{n}\!-\!R_{p} (\rm fm)$ \\
 \hline
 \hline\rule{0pt}{2.25ex} 
 \!\!FSUGarnet       & 7.892 &  5.493 & 0.0232 & 0.1612 \\ 
     FSUGarnet+$(f_{\rm v}\!>\!0)$  &  7.569 &  5.554 & 0.0235 & 0.1631  \\
     FSUGarnet+$(f_{\rm v}\!<\!0)$  &  8.242 &  5.429 & 0.0232 & 0.1622  \\ 
 \hline\rule{0pt}{2.25ex} 
     Experiment   &  7.867 &    5.501(1) & 0.041(13) & 0.283(71) \\
 \hline                                                                                                 
 \hline\rule{0pt}{2.25ex}   
\!\!Model (${}^{48}$Ca)& $B/A (\rm MeV)$ & $R_{\rm ch} (\rm fm)$ & $F_{\rm ch}\!-\!F_{\rm wk}$ & $R_{n}\!-\!R_{p} (\rm fm)$ \\
 \hline
 \hline\rule{0pt}{2.25ex} 
 \!\!FSUGarnet       &  8.621 &  3.428 & 0.0437 & 0.1665 \\ 
     FSUGarnet+$(f_{\rm v}\!>\!0)$  &  8.029 &  3.511 & 0.0430 & 0.1862  \\
     FSUGarnet+$(f_{\rm v}\!<\!0)$  &  9.316 &  3.339 & 0.0441 & 0.1449  \\ 
 \hline\rule{0pt}{2.25ex} 
     Experiment     & 8.667 & 3.477(2) & 0.0277(55) & 0.121(35) \\
 \hline                                                                                                 
\end{tabular}
\caption{Ground-state observables for $^{208}$Pb and $^{48}$Ca as predicted by the original FSUGarnet model\,\cite{Chen:2014mza}
              and the FSUGarnet model supplemented by a tensor-coupled $\omega$-meson, with $f_{\rm v}\!=\!\pm5$. No attempt has 
              been made at this point to re-calibrate the model parameters.}
\label{Table2}
\end{center}
\end{table}
%%%%%%%%%%%%%%%%%%%%

%%%%%%%%%%%%%%%%%%%%%%%%%%%%%%%%%%%%%%%%%%%%%%%%%%%%%%%%%%%%%%%%%%%
%%%%%%%%%%%%%%%%%%%%%%%%%%%%%%%%%%%%%%%%%%%%%%%%%%%%%%%%%%%%%%%%%%%
% FIXED ABOVE

\begin{center}
\begin{figure}[H]
\centering
\includegraphics[width=0.48\textwidth]{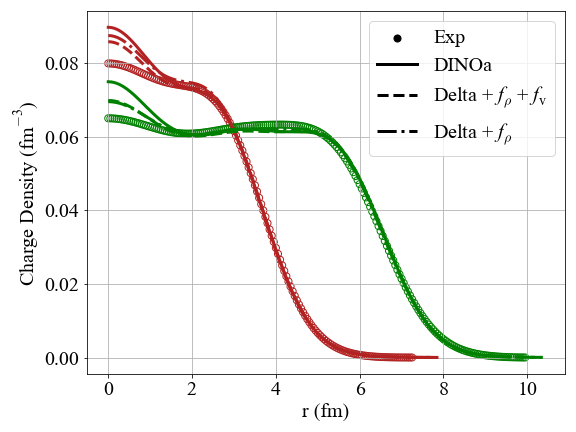}
\caption{Charge density distributions for $^{48}$Ca (red) and $^{208}$Pb (green). The effect of adding the two tensor interactions 
is seen to reduce the fluctuations in the nuclear interior that is characteristic of the DINO models.}
\label{Fig2}
\end{figure}
\end{center}

Given that large and cancelling $\sigma$- and $\omega$-meson fields are the hallmark of covariant EDFs, any modification to the 
$\omega$-meson field is expected to strongly impact the nuclear binding energy. This is clearly the case for both nuclei, but more 
so for $^{48}$Ca where the spread can exceed 1\,MeV per nucleon. Moreover, due to their difference in the surface-to-volume ratio,
the neutron skin thickness in $^{48}$Ca changes significantly, whereas it remains fairly constant in the case of $^{208}$Pb. Although,
encouraging, without a proper re-tuning of parameters these results are deceptive. In particular, since the $\omega$-meson generates
an isoscalar mean field, the spatial distribution of both neutron and protons shift in the same direction. So whereas we observe a differential
change in the neutron skin thickness of $^{48}$Ca relative to $^{208}$Pb, demanding agreement with the experimental binding energies
and charge radii is likely to wash out the desired differential effect. Nevertheless, the inclusion of $f_{\rm v}$ into the functional seems to 
improve the overall performance of the model. We now demonstrate this by re-calibrating the DINOa model after adding the two tensor 
couplings. This results in a functional that improves the interior charge densities stemming from the large delta coupling, while keeping 
$F_{\rm Wskin}$ mostly intact. The charge density distributions are displayed in Fig.\ref{Fig2} and the various nuclear observables are 
listed in Table\,\ref{Table3}. 

%%%%%%%%
%%%%%%%%

%%%%%%%%  Table III  %%%%%%%
\begin{table}[h]
\begin{center}
\begin{tabular}{|l||c|c|c|c|c|}
 \hline\rule{0pt}{2.25ex}   
 \!\!Model (${}^{208}$Pb)& $B/A (\rm MeV)$ & $R_{\rm ch} (\rm fm)$ & $F_{\rm ch}\!-\!F_{\rm wk}$ & $R_{n}\!-\!R_{p} (\rm fm)$ \\
 \hline
 \hline\rule{0pt}{2.25ex} 
 \!\!DINOa       & 7.864 &  5.508 & 0.0264 & 0.1753 \\ 
     DINOa$+\!f_\rho\!+\!f_{\rm v}$  &  7.871 &  5.510 & 0.0241 & 0.1616  \\ 
 \hline\rule{0pt}{2.25ex} 
     Experiment   &  7.867 &    5.501(1) & 0.041(13) & 0.283(71) \\
 \hline                                                                                                 
 \hline\rule{0pt}{2.25ex}   
\!\!Model (${}^{48}$Ca)& $B/A (\rm MeV)$ & $R_{\rm ch} (\rm fm)$ & $F_{\rm ch}\!-\!F_{\rm wk}$ & $R_{n}\!-\!R_{p} (\rm fm)$ \\
 \hline
 \hline\rule{0pt}{2.25ex} 
 \!\!DINOa       &  8.674 &  3.468 & 0.0334 & 0.1002 \\ 
     DINOa$+\!f_\rho\!+\!f_{\rm v}$  &  8.674 &  3.463 & 0.0344 & 0.1063  \\ 
 \hline\rule{0pt}{2.25ex} 
     Experiment     & 8.667 & 3.477(2) & 0.0277(55) & 0.121(35) \\
 \hline                                                                                                 
\end{tabular}
\caption{Ground-state observables for $^{208}$Pb and $^{48}$Ca as predicted by the original DINOa model\,\cite{Reed:2023cap}  
              and a re-calibrated DINOa model after including the two tensor interactions.} 
\label{Table3}
\end{center}
\end{table}
%%%%%%%%%%%%%%%%%%%%

It is important to note that after re-calibration, the addition of an $\omega$-meson tensor coupling does not always improve 
the agreement with experiment. There is some evidence suggesting that such a term has a strong impact in the determination
of the compressibility $K_0$ of the model \cite{Rufa:1988zz}. This could be especially important for models constrained
by experimental information on the Giant Monopole Resonance (GMR). Because information on the GMR was not included in our calibration procedure, our models tend to favor a relatively large $K_0$. For this reason the compressibility $K_0$ is capped at 250 MeV, a value within the suggested experimental limits\,\cite{Garg:2018uam}. Moreover, we
retain the $\omega$-tensor interaction as a means to improve the performance of the model. Later on, we will be able to
reliably asses the importance of such a term by invoking the full power of reduced basis methods. 

We will now divert our attention to the new scalar mixing term $\Lambda_{\rm s}$, which couples the isoscalar $\sigma$- and 
isovector $\delta$-meson in a manner analogous to how the $\Lambda_{\rm v}$ term acts on the vector sector. 

%%%%%%%%%%%%%%%%%%%%%%%%%%%%%%%%%%%%%%%%%%%%%%%%%%%%%%%%%%%%%%%%%%%
%%%%%%%%%%%%%%%%%%%%%%%%%%%%%%%%%%%%%%%%%%%%%%%%%%%%%%%%%%%%%%%%%%%
% FIXED ABOVE
\subsection{Effect of scalar mixing $\Lambda_s$}
\label{Sec:scalar mixing}

Earlier covariant EDFs with only one isovector parameter (${g_\rho}^{2}/m_{\rho}^{2}$) fitted to the symmetry energy at 
saturation density are naturally stiff\,\cite{Lalazissis:1996rd}. In an effort to tune the slope of the symmetry energy $L$---and 
therefore adjust the neutron skin thickness of ${}^{208}$Pb---a meson-mixing parameter in the vector sector ($\Lambda_{\rm v}$) 
was introduced\,\cite{Horowitz:2000xj}. Having now introduced the $\delta$-meson into the Lagrangian density, it is only natural to 
include a similar scalar mixing term $\Lambda_{\rm s}$ between the isoscalar $\sigma$-meson and the isovector $\delta$-meson. 
Such a scalar mixing term seems to play an important role in the EOS of neutron star matter as well as on the weak skin 
of $^{48}$Ca and $^{208}$Pb \cite{Li:2022okx,Miyatsu:2023lki}. 

One of the main drawbacks of models that are stiff above saturation density is the resulting large neutron star radii and high tidal 
deformabilities. Even the DINOa model with a fairly modest value for $L$ generates large tidal deformabilities because of the large 
and positive value for $K_{\rm sym}$. In an effort to soften the symmetry energy we now proceed to examine the effect of 
$\Lambda_{\rm s}$ on the DINOa model. Again, given that no attempt is made to re-calibrate the model parameters, we now
study separately the impact of both a positive and negative $\Lambda_{\rm s}$. 

It is important to note, however, that such meson mixing terms result in a re-definition of the effective meson mass\,\cite{Horowitz:2001ya}. 
In the particular case of a negative $\Lambda_{\rm v}$ or $\Lambda_{\rm s}$, there is a threshold density at which the meson 
mass becomes imaginary. Such a threshold density is higher for scalar mixing because of the large $\delta$-meson mass, so we proceed 
to explore the impact on the the DINOa model by arbitrarily setting the scalar mixing term to $\Lambda_{\rm s}\!=\!\pm 0.00025$. We 
underscore, however, that a larger and negative $\Lambda_{\rm s}$ may result in unphysical results once the effective $\delta$-meson
mass becomes negative. 

\begin{center}
\begin{figure}[H]
\centering
\includegraphics[width=0.45\textwidth]{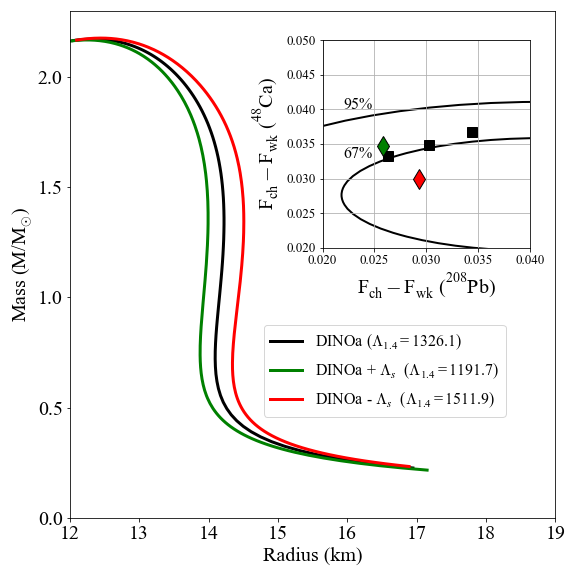}
\caption{Impact of the scalar mixing term on both neutron star radii and the weak-skin form factor of $^{48}$Ca and $^{208}$Pb (see inset
              on the upper right panel). The predictions of the three DINO models for the weak-skin form factors are shown along with the results 
              obtained after adding a positive (green) and negative (red) scalar mixing term of magnitude $\Lambda_{\rm s}\!=\!\pm 0.00025$. The 
              effects on stellar radii and the tidal deformability are displayed in the figure and the labels, respectively.}
\label{Fig3}
\end{figure}
\end{center}

We display in Fig.\ref{Fig3} mass radius relations as predicted by the original DINOa model, together with the predictions after adding the scalar 
mixing term. We note that for $\Lambda_{\rm s}\!>\!0$, both stellar radii and tidal deformability of a $1.4\,M_{\odot}$ neutron star are reduced relative 
to the original DINOa predictions; the effect, however, goes in the opposite direction for negative $\Lambda_{\rm s}$---exacerbating the undesired 
behavior of the original model. Yet, the effect is opposite in the case of the weak-skin form factor, shown on the inset in Fig.\ref{Fig3}. In this case 
the impact of a positive $\Lambda_{\rm s}$ is marginal, whereas a negative $\Lambda_{\rm s}$ brings the predictions well inside the 67\% confidence 
ellipse. In essence, a negative $\Lambda_{\rm s}$ brings the predictions into agreement with both PREX and CREX---but at the expense of increasing 
neutron star radii and tidal deformabilities. Our goal is then to use the new scalar mixing term to lower neutron star radii and tidal deformabilities, even 
if the impact on the weak skin is minimal. 

%%%%%%%
%%%%%%%

Although illuminating on their own, it is difficult to predict the combined effect of all the new terms that have been added to the model. In the next 
section we include all these new terms in an attempt to optimize a likelihood function containing finite nuclei observables. Once calibrated, we will
then examine the predictions for other relevant observables.

%%%%%%%%%%%%%%%%%%%%%%%%%%%%%%%%%%%%%%%%%%%%%%%%%%%%%%%%%%%%%%%%%%%
%%%%%%%%%%%%%%%%%%%%%%%%%%%%%%%%%%%%%%%%%%%%%%%%%%%%%%%%%%%%%%%%%%%
% FIXED ABOVE
\subsection{New Calibrated Functionals}

In this section we perform the calibration of a set of new functionals that relative to the DINO models include three additional terms with parameters
$f_{\rm v}, f_\rho$, and $\Lambda_{\rm s}$. For now, we set the quartic $\rho$-meson coupling $\xi$ to zero as its impact on finite nuclei observables is 
small\,\cite{Mueller:1996pm}. Later on, we will assess its its impact on the EOS at high densities. Although the quartic $\omega$-meson coupling $\zeta$ 
softens the EOS at high densities, it has been shown to have a negligible impact on the EOS in the vicinity of saturation density\,\cite{Mueller:1996pm}.
Yet, given its importance at high densities and its strong impact on the maximum neutron star mass\,\cite{Salinas:2023nci}, we fix the value of the quartic 
$\omega$-meson coupling to $\zeta\!=\!0.015$ in order to satisfy the current maximum-mass constraint from 
PSR J0740+6620\,\cite{Cromartie:2019kug,Fonseca:2021wxt}. Lastly, the new scalar mixing coupling $\Lambda_{\rm s}$ is constrained to avoid that 
any of the isovector meson masses become imaginary while $\Lambda_{\rm v}$ is being varied. The rest of the parameters, namely, $m_{\rm s}, 
g_{\rm s}^2, g_{\rm v}^2, g_\delta^2, g_\rho^2, \kappa, \lambda, \Lambda_{\rm v}, f_{\rm v},$ and $f_\rho$ are allowed to vary and fit to the binding energies and charge radii of $^{40}$Ca, $^{48}$Ca, and $^{208}$Pb. The weak skin form factors are also fitted to their experimental values. Besides binding 
energies, charge radii, and weak skin form factors, we supplement the set of observables used for the calibration by including an approximate value for the charge density at 
the origin in order to mitigate the large density fluctuations in the nuclear interior.

%% FIX BELOW
As mentioned earlier, the calibration of the models parameters is limited to the optimization of a likelihood function without any attempt to 
assign theoretical uncertainties. Such a challenging task will be performed in the near future by bringing to bear the full power of reduced basis 
methods. The newly calibrated models will henceforth be referred  as FSU-TD0, FSU-TD1, and 
FSU-TD2. The FSU-TD0 model differs from the other two---and indeed from the DINO models---in that the $\delta$-meson coupling is set to zero 
(i.e., $g_{\delta}\!=\!0$). The reason behind this choice is to isolate the impact of the delta coupling on a functional with tensor interactions. It also suggests that adding just a single tensor coupling to the FSU functional can offer drastic improvements to the densities (see Fig.\ref{Fig0}), neutron star radii, and tidal deformabilities. The other two models, FSU-TD1 and FSU-TD2, differ in that the former is calibrated 
assuming a negative scalar mixing term whereas the latter assumes $\Lambda_{\rm s}\!>\!0$. Model parameters and associated predictions for the bulk
properties of infinite nuclear matter are listed in Table\,\ref{Table4} and Table\,\ref{Table5}, respectively.

%%%%%  Table A  %%%%%
\onecolumngrid
\begin{center}
\begin{table}[h]
\begin{tabular}{|l||c|c|c|c|c|c|c|c|c|c|c|c|c|c|}
\hline\rule{0pt}{2.5ex}   
\!Model   &  $m_{\rm s}$  & $g_{\rm s}^2$  &  $g_{\rm v}^2$  &  $g_{\rho}^2$  & $g_\delta^2$ &  $\kappa$ &  $\lambda$  
               &  $\zeta$    &   $\Lambda_{\rm v}$ & $\Lambda_s$ & $f_v$ & $f_\rho$ \\
\hline
\hline
%DINOa+$f_v + f_\rho$   & 516.547 & 114.885 & 180.593 & 633.520 & 801.615 & 3.21116 & -0.00670425 & 0.015 & 0.0 & 0.00213536 & 0.0 & 2.32958 & -5.489 \\
%FSU-TD0 & 504.887 & 121.876 & 204.744 & 308.933 & 0.0 & 2.95569 & -0.00667997 & 0.015 & 0.0 & 0.0404966 & 0.0 & 0.0 & -80.0 \\
%FSU-TD1 + ($\xi$) & 523.350 & 137.752 & 216.245 & 440.198 & 462.522 & 3.00645 & -0.0073519 & 0.015 & (0.02) & 0.00285816 & -0.0005 & 5.08122 & -23.2207 \\
%FSU-TD2 + ($\xi$) & 520.948 & 128.648 & 202.611 & 488.902 & 504.015 & 3.02088 & -0.00709203 & 0.015 & (0.023) & 0.00428229 & 0.0005 & 3.6763 & -31.332 \\
FSU-TD0 & 504.887 & 121.876 & 204.744 & 308.933 & 000.000 & 2.95569 & -0.00667997 & 0.015 & 0.04049660 & \;0.0000  & 0.0000   & -80.0000 \\
FSU-TD1 & 523.350 & 137.752 & 216.245 & 440.198 & 462.522 & 3.00645 & -0.00735190 & 0.015 & 0.00285816 & -0.0005    &  5.0812 & -23.2207 \\
FSU-TD2 & 520.948 & 128.648 & 202.611 & 488.902 & 504.015 & 3.02088 & -0.00709203 & 0.015 & 0.00428229 & \;0.0005  & 3.6763 & -31.3320 \\
\hline
\end{tabular}
\caption{Model parameters for the newly optimized FSU-TD covariant EDFs. The parameter $\kappa$ and the the scalar-meson mass are given in MeV. 
              The other meson masses are fixed at $m_{\rm v}\!=\!782.5\,{\rm MeV}$, $m_{\rho}\!=\!763.0\,{\rm MeV}$, and $m_{\delta}\!=\!980.0\,{\rm MeV}$,
              respectively. The nucleon mass has been fixed at $M\!=\!939\,{\rm MeV}$.}
\label{Table4}
\end{table}
\end{center}
\twocolumngrid
%%%%%%%%%%%%%%%%%%%%

%%%%%  Table 6  %%%%%
%%%
\onecolumngrid
\begin{center}
\begin{table}[h]
\begin{tabular}{|l||c|c|c|c|c|c|c|c|c|c|}
\hline\rule{0pt}{2.5ex}   
\!Model   &  $\epszero$  & $\rhozero$  &  $M^*$  &  $K_0$  & $\tilde{J}$ &  $J$ & $L$ &  $K_{\rm sym}$ \\
\hline
\hline
%DINOa+$f_v + f_\rho$ & -16.1924 & 1.30594 & 0.583453 & 250.0 & 27.0 & 31.0678 & 50.0 & 506.0 \\
FSU-TD0 & -16.145 & 0.150 & 0.546 & 250.0 & 27.0&  31.466 & 50.0 & 138.137 \\
FSU-TD1 & -16.113 & 0.151 & 0.524 & 250.0 & 27.0 & 31.047 & 50.0 & 500.052 \\
FSU-TD2 & -16.155 & 0.152 & 0.544 & 250.0 & 27.0 & 30.649 & 50.0 & 500.069\\
\hline
\end{tabular}
\caption{Bulk properties of infinite nuclear matter [Eq.(\ref{EandS})] as predicted by the newly optimized FSU-TD 
covariant EDFs. All properties are defined at saturation density $\rhozero$, except for $\tilde{J}$ that is defined 
as the value of the symmetry energy at a density of two thirds of $\rhozero$. All quantities are in units of MeV,
except $\rhozero$ which is given in ${\rm fm}^{-3}$.}
\label{Table5}
\end{table}
\end{center}
\twocolumngrid
%%%%%%%%%%%%%%%%%%%%

%%%%%%%%%
\onecolumngrid
\begin{center}
\begin{figure}[h!]
\centering
\includegraphics[width=0.9\textwidth]{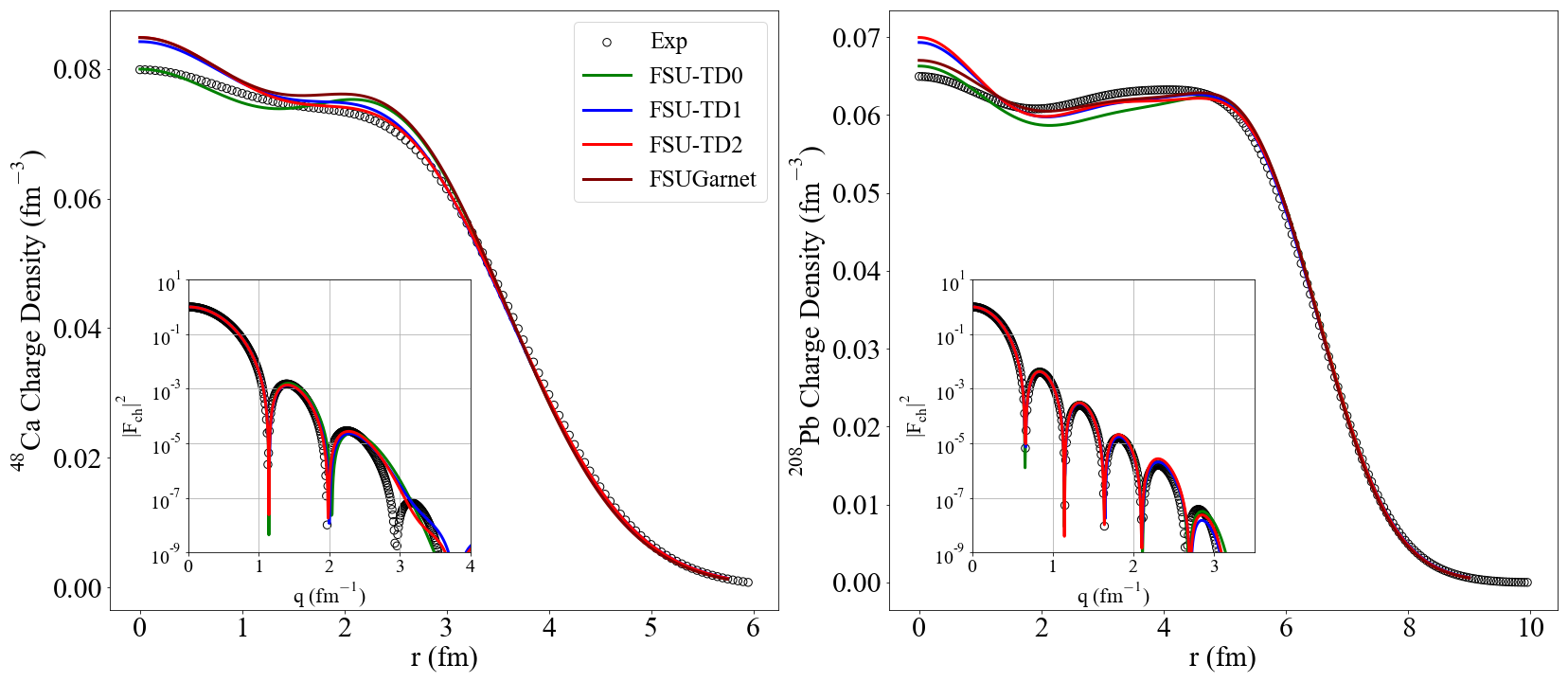}
\caption{Charge density distributions for $^{48}$Ca (left) and $^{208}$Pb (right) as predicted by the three new FSU-TD models, alongside the predictions
from FSUGarnet\,\cite{Chen:2014mza}. Shown in the inset is the square of the charge form factor $|F_{\rm ch}|^2(q)$ as a function of the momentum transfer.
The experimental charge form factors and corresponding spatial densities are obtained from Ref.\,\cite{DeJager:1987qc}.}
\label{Fig4}
\end{figure}
\end{center}
\twocolumngrid
%%%%%%%%%

%%%%%%%%  Table VI  %%%%%%%
\begin{table}
\begin{center}
\begin{tabular}{|l||c|c|c|c|c|}
 \hline\rule{0pt}{2.25ex}   
 \!\!Model (${}^{208}$Pb)& $B/A (\rm MeV)$ & $R_{\rm ch} (\rm fm)$ & $F_{\rm ch}\!-\!F_{\rm wk}$ & $R_{n}\!-\!R_{p} (\rm fm)$ \\
 \hline
 \hline\rule{0pt}{2.25ex} 
 \!\!FSU-TD0  & 7.871 & 5.523 & 0.0186 & 0.1322 \\ 
     FSU-TD1  & 7.871 & 5.511 & 0.0239 & 0.1618  \\ 
     FSU-TD2  & 7.869 & 5.510 & 0.0210 & 0.1428 \\ 
 \hline\rule{0pt}{2.25ex} 
     Experiment   &  7.867 &    5.501(1) & 0.041(13) & 0.283(71) \\
 \hline                                                                                                 
 \hline\rule{0pt}{2.25ex}   
\!\!Model (${}^{48}$Ca)& $B/A (\rm MeV)$ & $R_{\rm ch} (\rm fm)$ & $F_{\rm ch}\!-\!F_{\rm wk}$ & $R_{n}\!-\!R_{p} (\rm fm)$ \\
 \hline
 \hline\rule{0pt}{2.25ex} 
  \!\!FSU-TD0  & 8.661 & 3.446 & 0.0366 & 0.1287 \\  
       FSU-TD1 & 8.673 & 3.463 & 0.0335 & 0.1022  \\ 
       FSU-TD2 & 8.669 & 3.464 & 0.0319 & 0.0955 \\ 
 \hline\rule{0pt}{2.25ex} 
     Experiment     & 8.667 & 3.477(2) & 0.0277(55) & 0.121(35) \\
 \hline                                                
\end{tabular}
\caption{Ground-state observables for $^{208}$Pb and $^{48}$Ca as predicted by the newly calibrated FSU-TD models. The weak skin form factors $F_{\rm ch}\!-\!F_{\rm wk}$ for $^{48}$Ca and $^{208}$Pb include spin-orbit corrections\,\cite{Horowitz:2012we} and were computed at the momentum transfers of relevance to CREX ($q=0.8733\,{\rm fm}^{-1}$) and PREX  ($q=0.3977\,{\rm fm}^{-1}$).}
\label{Table6}
\end{center}
\end{table}
%%%%%%%%%%%%%%%%%%%%

\begin{center}
\begin{figure}
\centering
\includegraphics[width=0.99\textwidth]{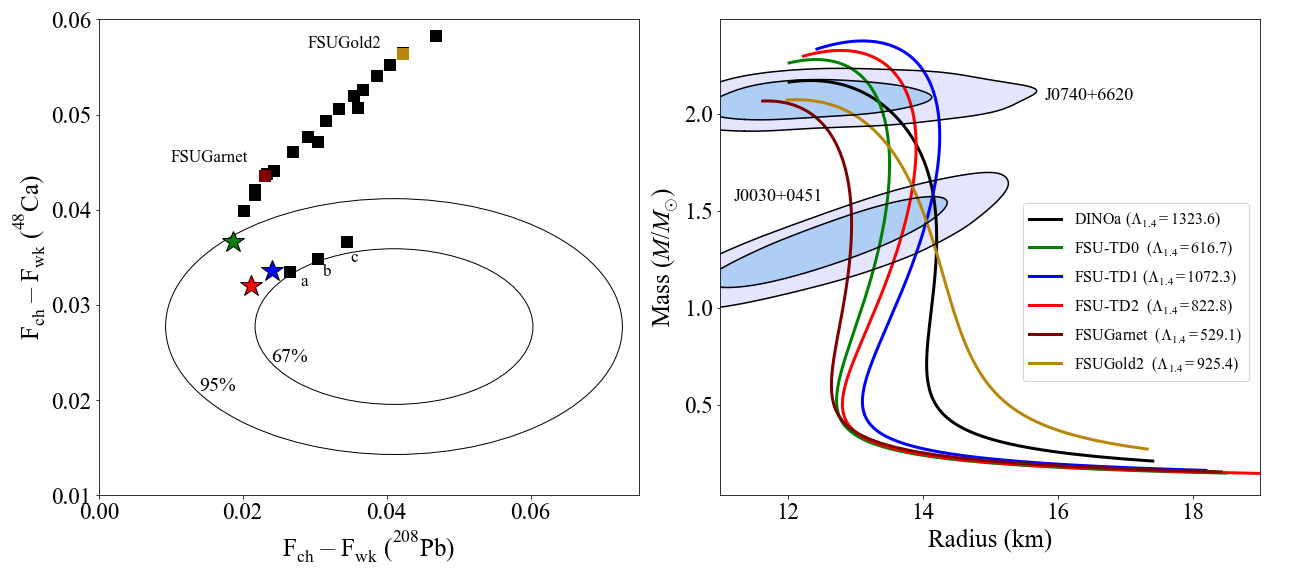}
\caption{Weak skin form factor $F_{\rm ch}\!-\!F_{\rm wk}$ for the new class of FSU-TD models displayed with stars in the left panel alongside other 
RMF models, such as FSUGarnet and FSUGold2. Shown together with the three DINO models labeled a,b,c are predictions from FSU-TD0 (green), 
FSU-TD1 (blue), and FSU-TD2 (red). The right panel shows the mass-radius relations as predicted by the various models and include in parenthesis,
the tidal deformabilities for a $M\!=\!1.4\,M_{\odot}$ neutron star. Also included are the 68$\%$ and 95$\%$ confidence intervals for the two NICER
sources J0030 and J0740.}
\label{Fig5}
\end{figure}
\end{center}

To demonstrate the fidelity of these new models, we display in Fig.\ref{Fig4} charge density distributions for $^{48}$Ca and $^{208}$Pb. The newly
calibrated FSU-TD models eliminate the large density oscillations in the nuclear interior displayed in Fig.\ref{Fig1}, without compromising the success 
of the models in reproducing binding energies and charge radii, as indicated in Table\,\ref{Table6}. For reference, we also show predictions from the 
FSUGarnet model\,\cite{Chen:2014mza} together with the experimental data of Ref.\cite{DeJager:1987qc}. Also shown as insets in the figure are the 
corresponding charge form factors normalized to $F_{\rm ch}(q\!=\!0)\!=\!1$. The curvature at the origin is proportional to the mean square radius of 
the charge distribution and is well reproduced by all the models. This is hardly surprising given that the charge radius is included in the calibration 
of the model parameters. However, the agreement with experiment extends up to at least $q\!=\!2\,{\rm fm}^{-1}$, suggesting that these class of 
covariant EDFs are robust to about such a momentum-transfer range. Beyond a few diffraction minima the agreement with experiment is lost, since 
short-range correlations and other effects that go beyond mean-field descriptions are important in accounting for the large-$q$ behavior of the form 
factor. We note that eliminating the large density fluctuations in the nuclear interior is mainly due to the inclusion of the tensor-coupled-$\rho$ interaction, 
which enables one to reduce the very large isovector ($g_{\delta}$ and $g_{\rho}$) coupling constants characteristic of the DINO models, while still 
generating weak skin form factors that fall within the 95\% confidence ellipse; see left-hand panel in Fig.\ref{Fig5}. 

%%%%%%%%
\onecolumngrid
\twocolumngrid
%%%%%%%%
While the newly calibrated DINO and FSU-TD models were developed with the aim of reconciling the CREX-PREX tension, it is imperative to achieve 
this goal without compromising the success of earlier models in reproducing the properties of both finite nuclei and neutron stars. To this end, we now display
neutron star predictions from the new models, focusing primarily on stellar radii and tidal deformabilities. A critical factor in mitigating the large tidal 
deformabilities predicted by the DINO models is the addition of a positive scalar coupling $\Lambda_{\rm s}$ to the Lagrangian density. In this context, 
the FSU-TD2 model stands out as unique, nearly falling within the 67\% CREX/PREX confidence ellipse, demonstrating good agreement with electron 
scattering experiments, and predicting stellar radii consistent with NICER, and tidal deformabilities within $1.6\,\sigma$ of the recommended value by
the LIGO-Virgo collaboration\,\cite{Abbott:2018exr}. Instead, the FSU-TD1 functional---with a negative $\Lambda_{\rm s}$---offers only marginal 
improvement in describing charge densities and weak skin form factors relative to FSU-TD2, but at a significant cost in reproducing stellar properties.
This behavior is depicted in the two panels in Fig.\ref{Fig5}, where the CREX-PREX confidence ellipse is displayed alongside neutron star predictions 
that are compared against NICER's 68\% and 95\% confidence contours for both J0030 and J0740. 

\begin{center}
\begin{figure}[H]
\centering
\includegraphics[width=0.99\textwidth]{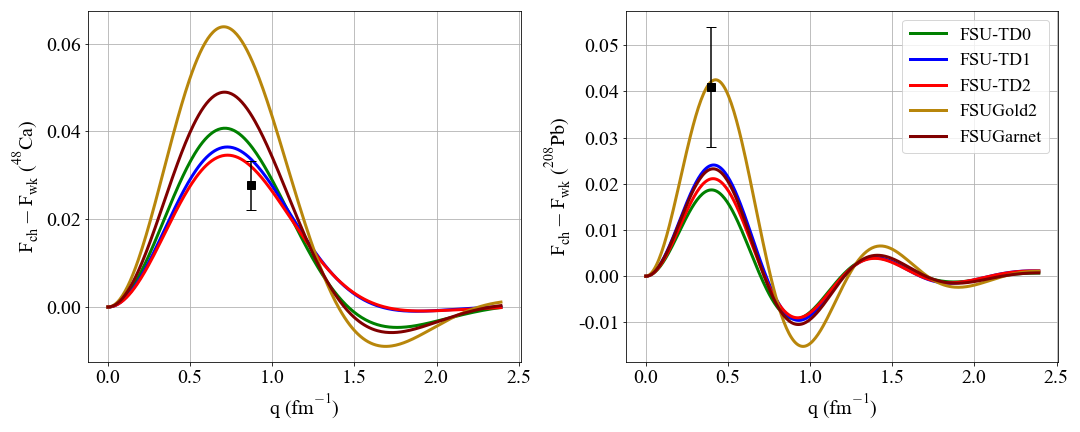}
\caption{Weak skin form factor $F_{\rm ch}\!-\!F_{\rm wk}$ as a function of momentum transfer $q$ for the new class of FSU-TD models. Also shown
with their associated colors are predictions from the FSUGold2\,\cite{Chen:2014sca} and FSUGarnet\,\cite{Chen:2014mza} models, alongside the CREX 
and PREX results. Note that while FSUGold2 pins down the PREX result, it dramatically overestimates the CREX result. In contrast, the softer FSUGarnet 
model shows some improvement for CREX but underestimates the PREX result.}
\label{Fig6}
\end{figure}
\end{center}

Particularly interesting is to explore the entire momentum-transfer dependence of the weak skin form factor\,\cite{Thiel:2019tkm}.
This is shown in Fig.\ref{Fig6} for the three new models alongside older predictions from FSUGold2\,\cite{Chen:2014sca} and FSUGarnet\,\cite{Chen:2014mza}.
The curvature at the origin is proportional to the weak skin defined in Eq.(\ref{Fwskin}). Also shown are the experimental measurements at the
relevant momentum transfers of $q=0.8733\,{\rm fm}^{-1}$ for CREX and $q=0.3977\,{\rm fm}^{-1}$ for PREX. It is clear from the figure that while the 
extraction of the weak skin involves some model dependence, $F_{\rm Wskin}$ is a genuine experimental observable. This figure---together with 
Fig.\ref{Fig5}---encapsulates the CREX-PREX dilemma: none of the large set of covariant EDFs can simultaneously reproduce the CREX and
PREX results at the 1$\sigma$ level. Yet as mentioned earlier, no theoretical model can convincingly resolve the CREX-PREX tension.
%%%
\onecolumngrid
\twocolumngrid

%%%%%%%%%%%%%%%%%%%%%%%%%%%%%%%%%%%%%%%%%
%%%%%%%%%%%%%%%%%%%%%%%%%%%%%%%%%%%%%%%%%%
%Fixed Above

\subsection{$\chi$EFT and the quartic coupling $\xi$}

We now proceed to address the role of the quartic $\rho$-meson coupling $\xi$. As argued by M\"uller and Serot in Ref.\cite{Mueller:1996pm}, 
the effect of $\xi$ at normal nuclear density is virtually imperceptible, but it may become important at high densities, especially when the proton fraction 
becomes very small. There they argue that the only significant nonlinearity at high density is due to the quartic, isoscalar vector interaction $\zeta$. 
However, we find that the inclusion of the quartic, isovector term $\xi$ becomes important when comparing the new models against predictions from 
Chiral Effective Field Theory ($\chi$EFT) for the EOS of Pure Neutron Matter (PNM)\,\cite{Drischler:2021kxf}. 

\begin{center}
\begin{figure}[h]
\centering
\includegraphics[width=0.40\textwidth]{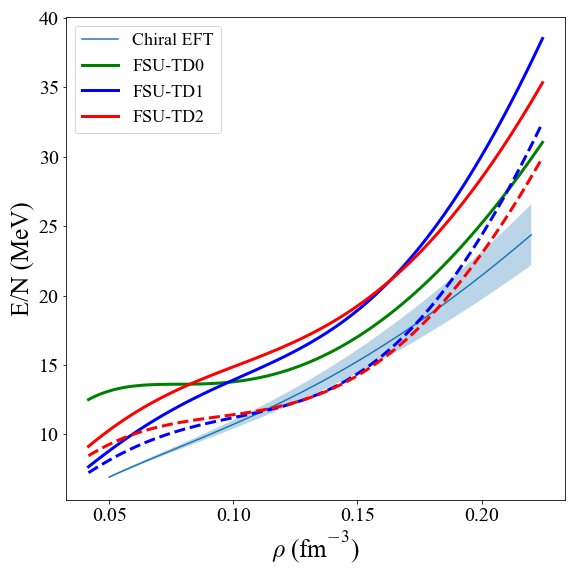}
\caption{The energy per neutron as a function of density. Results from $\chi$EFT\,\cite{Drischler:2021kxf} are shown in light blue with their 
associated uncertainty bands. The predictions of the three FSU-TD models are given with (dashed lines) and without (solid lines) the $\xi$ 
term. The $\xi$ coupling constant has no noticeable impact on the predictions of the FSU-TD0 model, so it is therefore omitted.}
\label{Fig7}
\end{figure}
\end{center}

Even for the FSU-TD0 model with relatively moderate values for the slope and curvature of the symmetry energy, namely $L\!=\!50\,{\rm MeV}$ and 
$K_{\text{sym}}\!=\!138\,{\rm MeV}$, the predicted behavior for the EOS of PNM deviates significantly from the one suggested by $\chi$EFT; see 
Fig.\ref{Fig7}. Such large deviations relative to $\chi$EFT especially at low densities is concerning since it is in this region where the $\chi$EFT 
uncertainties are extremely small. To mitigate this problem we invoke the quartic, isovector term $\xi$. Because the imperceptible impact of $\xi$ 
on the properties of finite nuclei, the model parameters listed in Table\,\ref{Table4} remain unchanged. This is because the proton fraction of the 
finite nuclei that define the fitting protocol is relatively large. Hence, without any further calibration of parameters, we simply estimate the value of 
$\xi$ in an effort to reduce the discrepancy with $\chi$EFT. 

We set the values of quartic, isovector term to $\xi\!=\!0.020$, and $\xi\!=\!0.023$ for FSU-TD1 and FSU-TD2, respectively. Note that we have found 
no significant improvement on the EOS of PNM for the FSU-TD0 model, so we simply set the value of $\xi$ to zero. The impact of adding $\xi$ on 
the other two models is depicted by the dashed lines in Figure \ref{Fig7}. Although clearly not perfect, the addition of the quartic, isovector term 
improves the agreement with $\chi$EFT. Moreover, because the influence of $\xi$ is strong in the vicinity and of nuclear saturation density, we are 
confident that the discrepancy with $\chi$EFT can be remedied without much difficulty, once suitable priors are included in the calibration of the 
model. Indeed, it was shown in Ref.\,\cite{Salinas:2023nci} that such a region is mostly sensitive to $\epszero$, $J$, and $L$, bulk parameters of 
infinite nuclear matter that are well constrained by existing nuclear data.

%%%%%%%%%%%%%%%%%%%%%%%%%%%%%%%%%%%%%%%%%%%%%%%%%%%%%%
%%%%%%%%%%%%%%%%%%%%%%%%%%%%%%%%%%%%%%%%%%%%%%%%%%%
%GOOD ABOVE

%%%%%%%
%%%%%%%

\begin{center}
\begin{figure}[h]
\centering
\includegraphics[width=0.48\textwidth]{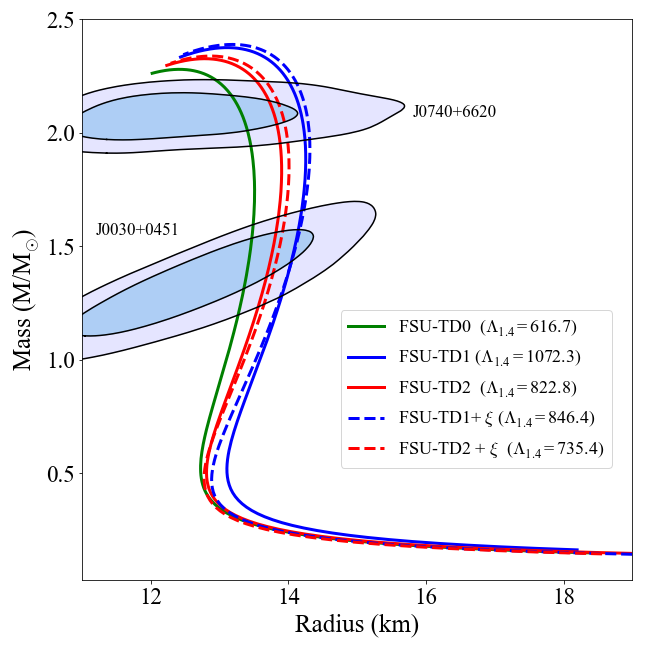}
\caption{Mass-radius relations for the FSU-TD models (solid lines) and the FSU-TD models supplemented by the addition of the quartic, isovector term $\xi$ (dashed lines). Also shown are NICER's $68\%$ and $95\%$ confidence contours together with predictions for the tidal deformabilities of a $M\!=\!1.4\,M_{\odot}$ neutron star (see legends).}
\label{Fig8}
\end{figure}
\end{center}

Given that the addition of $\xi$ affects the EOS of neutron star matter at and above saturation density, we also see improvements
in the description of neutron star observables, particularly in the case of the tidal deformability for both FSU-TD1, and FSU-TD2. 
Yet, we underscore that unlike the $\zeta$ term which primarily affects the high density region of the EOS, the quartic, isovector 
parameter $\xi$ is relevant even at intermediate densities given that neutron star matter becomes more symmetric with increasing 
density. As such, $\xi$ softens the EOS in a density region of relevance to intermediate-mass neutron stars, such as in the binary 
system GW170817\,\cite{Abbott:PRL2017}. This argument is validated in Fig.\ref{Fig8} where we observe a significant decrease in the 
tidal deformability of a $M\!=\!1.4\,M_{\odot}$ neutron star---not necessarily because of a large reduction in the stellar radii, 
but rather due to a reduction in the second Love number which is sensitive to the EOS. Indeed, as documented in Table\,\ref{Table7}, 
the inclusion of $\xi$ results in the following favorable changes to the tidal deformability: 
$\Lambda_{1.4}\!=\!1072\rightarrow846$ for FSU-TD1 and $\Lambda_{1.4}\!=\!823\rightarrow735$ for FSU-TD2, respectively. A future 
sensitivity study of $\xi$ on both the Love number and the tidal deformability could be helpful in understanding its impact on the 
EOS\,\cite{Piekarewicz:2018sgy}. Also shown in Table\,\ref{Table7} are the predicted maximum mass supported by the EOS, the core-crust 
transition density, and threshold density and mass for the direct Urca process.  

%%%%%%%%
%%%%%%%%

%%%%%  Table 7 %%%%%
%%%%%%%%%%%%%%%
\onecolumngrid
\begin{center}
\begin{table}[h]
\begin{tabular}{|l||c|c|c|c|c|c|c|}
\hline\rule{0pt}{2.5ex}   
\!Model   &  $R_{1.4}$ (km)  & $\Lambda_{1.4}$  & $k_{2}(1.4)$ & $M_{\text{max}}$ ($M_\odot$) &  $\rho_{\text{trans}}$ (fm$^{-3}$) & $\rho_{\text{Urca}}$ (fm$^{-3}$) &  $M_{\text{Urca}}$ (M$_\odot$) \\
\hline
\hline
FSU-TD0 &          13.41 & \;616.7  & 0.0808 & 2.279 & 0.0824 & 0.489 & 2.068 \\
FSU-TD1 &          14.01 & 1072.3  & 0.1119 & 2.375 & 0.1037 & 0.261  & 1.284 \\
FSU-TD2 &          13.71 & \;822.8 & 0.0959 & 2.326 & 0.1036 & 0.303 & 1.477 \\
FSU-TD1 + $\xi$ &  14.02 & \;846.4  & 0.0889 & 2.389 & 0.0950 & 0.298 & 1.641 \\
FSU-TD2 + $\xi$ &  13.79 & \;735.4  & 0.0831 & 2.338 & 0.0910 & 0.331 & 1.699 \\
\hline
\end{tabular}
\caption{Neutron star predictions for the FSU-TD models with and without the inclusion
of the quartic $\rho$-meson self interaction. Shown are the radius, dimensionless tidal 
deformability, and second Love number for a $1.4\,M_{\odot}$ neutron star. Also shown
are the maximum mass supported by the EOS, the core-crust transition density, and threshold
density and mass for the direct Urca process. The core-crust transition density is 
computed using the Thermodynamic Stability Method of Ref.\cite{Routray:2016yvp}.}
\label{Table7}
\end{table}
\end{center}
\twocolumngrid
%%%%%%%%%%%%%%%%%%%%

We conclude this section with a word of caution on developing insights on the symmetry energy and the EOS of neutron star matter from 
$\chi$EFT. As indicated in Eq.(\ref{EOS1}), the nuclear symmetry energy is strictly defined as 
%%%%%%%%%
\begin{equation}
  S(\rho) = \frac{1}{2}\!\left(\frac{\partial^{2} E(\rho,\alpha)}{\partial\alpha^{2}}\right)_{\!\!\alpha\!=\!0}.
 \label{SyE}
\end {equation}
%%%%%%%%%
Yet, for a large class of functionals the symmetry energy seems to be well reproduced by the so-called parabolic approximation, 
defined as the difference between the EOS of pure neutron matter and the corresponding one for symmetric nuclear matter, i.e., 
$S(\rho)\!=\!E_{\rm PNM}(\rho)\!-\!E_{\rm SNM}(\rho)$. It is this parabolic approximation that is often used to construct the
EOS of charge-neutral, neutron star matter in beta equilibrium\,\cite{Drischler:2020fvz}. Indeed, based on this approximation, 
$\chi$EFT approaches develop a framework in which the energy per particle of asymmetric nuclear matter is obtained 
from an interpolation between the EOS of symmetric nuclear matter and that of pure neutron matter\,\cite{Hebeler:2013nza}. 
It is this expression, together with the leptonic contribution to the EOS, that is then used to determine the proton fraction for 
neutron star matter.

\begin{center}
\begin{figure}[h]
\centering
\includegraphics[width=0.48\textwidth]{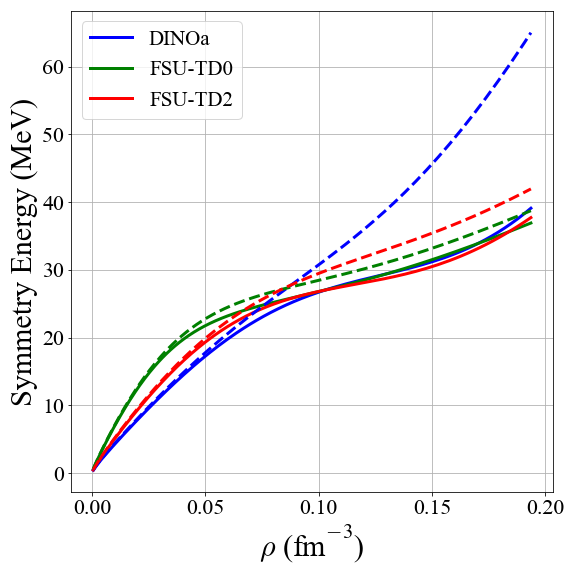}
\caption{The symmetry energy calculated for some of the new classes of models with enlarged isovector sectors. The result using 
the parabolic approximation (dashed) differs---in some instances significantly---from the result using the exact expression for the 
symmetry energy (solid) depending on the strength of the isovector coupling constants.}
\label{Fig9}
\end{figure}
\end{center}

However, we note that for the new class of covariant EDFs considered here, the parabolic approximation may be badly broken.
We display in Fig.\ref{Fig9} a comparison between the symmetry energy computed exactly (solid lines) as in Eq.(\ref{SyE}) 
and in the parabolic approximation (dashed lines). It is interesting to note that in the case of the DINOa model---with very large
Yukawa couplings in the isovector sector---the parabolic approximation already becomes invalid at low densities and is badly 
broken at saturation density. In the case of the FSU-TD models, the additional isovector coupling constants reduce the Yukawa 
$\delta$-meson coupling resulting in better agreement between the two prescriptions. Yet, we must conclude that while $\chi$EFT calculations
of pure neutron matter are enormously useful for the calibration of energy density functionals, relying on the parabolic approximation
to compute the equation of state of neutron star matter should be carefully scrutinized. 

%%%%%%%%
%%%%%%%%

%%%%%%%%%%%%%%%%%%%%%%%%
%FIXED ABOVE
%%%%%%%%%%%%%%%%%%%%%%%%%%%%%%%%%%%%%%%%%%%%%%%%%%
\section{Conclusions}
\label{Sec:Conclusions}

The highly successful PREX and CREX campaigns are challenging our most basic understanding of the nuclear symmetry 
energy. While the binding energy of nuclei with a moderate neutron excess determines the symmetry energy in the vicinity 
of nuclear saturation density, the neutron skin thickness has been identified as a robust proxy for the determination of its 
slope. Indeed, a large class of energy density functionals suggest a fairly strong correlation between the neutron skin thickness 
of $^{48}$Ca and $^{208}$Pb. This correlation now appears to be badly broken. Whereas PREX favors a large neutron skin 
thickness in $^{208}$Pb and thus a stiff symmetry energy, the significantly thinner neutron skin thickness in $^{48}$Ca
reported by CREX suggests instead a much softer symmetry energy. 

Given the relatively primitive nature of the isovector sector of most covariant energy density functionals, there have been several 
recent attempts to enlarge the isovector sector by including the isovector-scalar $\delta$-meson in search for a resolution to the 
CREX-PREX dilemma\,\cite{Li:2022okx,Thakur:2022dxb,Miyatsu:2023lki}. In this manuscript we investigate the impact of a significantly 
enlarged Lagrangian density on 
the properties of both finite nuclei and neutron stars. Relative to the Lagrangian density of Ref.\cite{Chen:2014sca} containing
only two isovector parameters, we have added: (a) the $\delta$-meson and its non-linear coupling to the isoscalar-scalar 
$\sigma$-meson, (b) tensor terms for both the isoscalar-vector $\omega$-meson and the isovector-vector $\rho$-meson, and 
(c) a quartic self-interacting coupling term for the $\rho$-meson. The addition of these new terms suggests a promising avenue
to resolve some outstanding problems concerning the structure of finite nuclei and neutron stars. 

We started our investigation by examining the impact of each new term on a few critical observables, such as the weak skin form 
factor and the mass-radius relation. We showed that the addition of tensor interactions plays a crucial role in reducing the large
oscillations in the charge density displayed by the DINO models in the nuclear interior, while preserving the agreement of the 
DINO models in mitigating the CREX-PREX discrepancy. Moreover, the inclusion of the mixed scalar coupling $\Lambda_{\rm s}$
together with the quartic $\rho$-meson self-interaction $\xi$ is shown to have a significant impact on the EOS at intermediate
and high densities. The addition of these two terms also alleviates some of the problems of the DINO models that arise because
of the large value of the curvature of the symmetry energy $K_{\rm sym}$ required to modify the symmetry energy in the vicinity
of the saturation density. Such large values for $K_{\rm sym}$ predict large stellar radii that are in conflict with the NICER results
and overestimate the tidal deformability reported by the LIGO-Virgo collaboration. To a large extent, these problems have been 
solved in the present work. Nevertheless, without a proper calibration procedure that provides model uncertainties and correlations 
between observables, the full impact of the new Lagrangian density is difficult to assess. Yet, we are optimistic that the significant 
advancements in employing reduced basis models for implementing Bayesian inference will aid us in achieving this goal.

\begin{acknowledgments}\vspace{-10pt}
We thank the Institute for Nuclear Theory at the University of Washington for its kind hospitality and stimulating research environment
during the program and workshop on ``Neutron Rich Matter on Heaven and Earth''. We also thank Prof. Anna Watts for many fruitful discussions on the NICER measurements and Prof. Nicholas Chamel and Nikolai Shchechilin for providing us with a crustal EOS. This material is based upon work supported 
by the U.S. Department of Energy Office of Science, Office of Nuclear Physics under Award Number DE-FG02-92ER40750. 
\end{acknowledgments}

%\bibliography{./ReferencesJP}

\end{document}